\documentclass[%
 reprint,
 preprintnumbers,
 amsmath,amssymb,
 aps,
]{revtex4-2}

\usepackage{graphicx}% Include figure files
\usepackage{dcolumn}% Align table columns on decimal point
\usepackage{bm}% bold math
\usepackage{hyperref}% add hypertext capabilities
\hypersetup{
	colorlinks=true,
	linkcolor=blue,
	filecolor=magenta,      
	urlcolor=blue,
	citecolor=magenta
}
\usepackage{amsmath,amssymb,mathtools}
\usepackage{graphics,graphicx,tikz,tensor}
\usetikzlibrary{arrows,decorations.pathmorphing,patterns}

\begin{document}

\preprint{NORDITA 2022-073}
\preprint{UUITP-46/22}

\title{Angular Momentum Loss Due to Tidal Effects\\in the Post-Minkowskian Expansion}% Force line breaks with \\

\author{Carlo Heissenberg$^{\ast,\dagger}$}%\email{carlo.heissenberg@physics.uu.se}
\affiliation{$^\ast$Department of Physics and Astronomy, Uppsala University, Box 516, SE-75237 Uppsala, Sweden\\
	$^\dagger$NORDITA, KTH Royal Institute of Technology and Stockholm University,
	Hannes Alfv\'ens väg 12, SE-11419, Stockholm, Sweden}

%\collaboration{}%\noaffiliation


\begin{abstract}
We calculate the tidal corrections to the loss of angular momentum in a two-body collision at leading Post-Minkowskian order from an amplitude-based approach. The eikonal operator allows us to efficiently combine elastic and inelastic amplitudes, and captures both the contributions due to genuine gravitational-wave emissions and those due to the static gravitational field. We calculate the former by harnessing powerful collider-physics techniques such as reverse unitarity, thereby reducing them to cut two-loop integrals, and cross-check the result by performing an independent calculation in the Post-Newtonian limit. For the latter, we can employ the results of \cite{DiVecchia:2022owy}, %arXiv:2203.11915
where static-field effects were calculated for generic gravitational scattering events using the leading soft graviton theorem. 
\end{abstract}

\maketitle

%\tableofcontents

\paragraph*{Introduction.} 
The steadily increasing sensitivity of gravitational-wave measurements challenges the state of the art of precision calculations for gravitational collisions \cite{Buonanno:2022pgc}.
In this context, scattering amplitudes have found fertile ground and contributed to advance the precision frontier in the Post-Minkowskian (PM) expansion, based on successive approximations labeled by powers of the Newton constant $G$ (see \footnote{We work in natural units commonly employed in the particle physics literature. Therefore, we measure velocities in units of the speed of light $c$, effectively setting $c=1$ and identifying length and time intervals. Similarly, we measure actions in units of the reduced Planck constant $\hbar$, so that $\hbar=1$. At variance with the General Relativity literature, instead, we keep explicit the dependence on the Newton constant $G$, which in natural units has the effective dimension $(\text{energy})^{-2}$. Of course, the appropriate powers of $c$ (and, if needed, of $\hbar$) can  always be reinstated via dimensional analysis.} for our conventions on the units of $G$, $c$ and $\hbar$) and valid for generic velocities \cite{Bjerrum-Bohr:2018xdl,Bern:2019nnu,Bern:2019crd,Bern:2021dqo,Bern:2021yeh}.
This synergy between general relativity and amplitude methods, and its recent successes highlight the importance of pushing these calculations to higher orders and of including all relevant physical effects, such as spin \cite{Arkani-Hamed:2017jhn,Vines:2017hyw,Guevara:2018wpp,Chung:2018kqs,Maybee:2019jus,Guevara:2019fsj,Arkani-Hamed:2019ymq,Johansson:2019dnu,Chung:2019duq,Damgaard:2019lfh,Bautista:2019evw,Aoude:2020onz,Chung:2020rrz,Bern:2020buy,Guevara:2020xjx,Kosmopoulos:2021zoq,Aoude:2021oqj,Bautista:2021wfy,Chiodaroli:2021eug,Haddad:2021znf,Chen:2021kxt,Aoude:2022trd,Bern:2022kto,Alessio:2022kwv,FebresCordero:2022jts,Cangemi:2022abk} and tidal corrections \cite{Cheung:2020sdj,Bern:2020uwk,Cheung:2020gbf,Aoude:2020ygw,AccettulliHuber:2020dal} that will be vital, in combination with numerical simulations, to provide accurate waveform models \cite{Buonanno:2022pgc}.
The measurement of effects due to tidal deformations \cite{Damour:1992qi,Damour:1993zn,Goldberger:2005cd,Hinderer:2007mb,Flanagan:2007ix,Damour:2009vw,Binnington:2009bb,Hinderer:2009ca,Kol:2011vg,Damour:2012yf,Favata:2013rwa}, in particular, may provide clues on the internal structure of neutron stars \cite{Baiotti:2016qnr}, 
on the nature of black holes \cite{Barack:2018yly} and on the possible existence of exotic
astrophysical objects \cite{Buonanno:2014aza,Cardoso:2019rvt,Baumann:2019ztm}.

Amplitudes provide a natural way to organize the $G$-expansion, based on the standard perturbative series where the double-copy \cite{Bern:2008qj,Bern:2010ue,Bern:2012uf,Bern:2017ucb,Bern:2018jmv,Bern:2019prr}, generalized unitarity \cite{Bern:1994zx,Bern:1994cg,Britto:2004nc} and gauge invariance offer powerful techniques for integrand construction. Resummation methods like the eikonal exponentiation \cite{Kabat:1992tb,Akhoury:2013yua,KoemansCollado:2019ggb,Cheung:2020gyp,DiVecchia:2020ymx,AccettulliHuber:2020oou,DiVecchia:2021ndb,DiVecchia:2021bdo,Heissenberg:2021tzo,Bjerrum-Bohr:2021vuf,Bjerrum-Bohr:2021din,Damgaard:2021ipf,Brandhuber:2021eyq,DiVecchia:2022nna}, effective-field-theory matching \cite{Goldberger:2004jt,Porto:2016pyg,Cheung:2018wkq,Cristofoli:2020uzm} or the KMOC framework \cite{Kosower:2018adc,Herrmann:2021lqe,Herrmann:2021tct,Cristofoli:2021vyo,Cristofoli:2021jas,Adamo:2022rmp,Adamo:2022qci} then provide the needed bridge between the quantum formulation and the classical PM regime of scattering at large impact parameter $b\gg Gm^\ast$, with $m^\ast$ the typical mass (or energy) scale of the colliding objects.
Moreover, techniques borrowed from collider physics like integration via differential equations \cite{Parra-Martinez:2020dzs,DiVecchia:2021bdo} and reverse unitarity \cite{Anastasiou:2002qz,Anastasiou:2002yz,Anastasiou:2003yy,Anastasiou:2015yha,Herrmann:2021lqe,Herrmann:2021tct} have recently proven very valuable when applied to the calculation of classical observables as well.
Such techniques and ideas have been also exploited in the context of quantum-field-theory-inspired worldline setups that efficiently encode the PM expansion \cite{Goldberger:2016iau,Kalin:2020mvi,Kalin:2020fhe,Kalin:2020lmz,Mogull:2020sak,Jakobsen:2021smu,Mougiakakos:2021ckm,Liu:2021zxr,Dlapa:2021npj,Jakobsen:2021lvp,Jakobsen:2021zvh,Riva:2021vnj,Dlapa:2021vgp,Jakobsen:2022fcj,Mougiakakos:2022sic,Jakobsen:2022psy,Kalin:2022hph,Dlapa:2022lmu,Jakobsen:2022zsx}.

In this paper we focus on dissipative effects induced by linear tidal deformations corresponding to mass or ``electric'' and current or ``magnetic'' quadrupole corrections. Combining eikonal operator \cite{Ciafaloni:2018uwe,Addazi:2019mjh,Damgaard:2021ipf,DiVecchia:2022nna,DiVecchia:2022owy,Cristofoli:2021vyo,DiVecchia:2022piu} and reverse unitarity, we first confirm the results of \cite{Mougiakakos:2022sic,Jakobsen:2022psy} for the radiated energy-momentum and then obtain a totally new prediction: the angular momentum lost due to tidal effects, thus completing the analysis of the Poincar\'e charges of the gravitational field to leading PM order, performed in \cite{Herrmann:2021lqe,Manohar:2022dea} for point particles, and initiated in \cite{Mougiakakos:2022sic,Jakobsen:2022psy} for tidal effects.
Two types of contributions are relevant for this calculation. The first is due to the emission of gravitational waves, described by superpositions of dynamically propagating gravitons. The second is due to static-field effects that are localized at the zero-frequency end of the graviton spectrum. Both fit naturally within our approach. 

Radiative contributions are calculated by recasting them as Fourier transforms of three-particle cuts, which can be in turn evaluated as cut two-loop integrals \cite{Herrmann:2021lqe,DiVecchia:2021bdo,Riva:2021vnj}.
Static contributions follow from the results of \cite{DiVecchia:2022owy}, where they were evaluated for generic processes exploiting the universality of the leading soft graviton theorem, supplemented by the tidal corrections to the impulse \cite{Cheung:2020sdj,Bern:2020uwk,Aoude:2020ygw}. 
	The fluxes of energy and angular momentum serve, in combination with the binding energies, as ingredients for building accurate waveform models \cite{Damour:2008yg,Antonelli:2019ytb,Khalil:2022ylj,Buonanno:2022pgc} that are crucial for gravitational-wave detection and analysis. For this reason, we also provide the analytic continuation of the result to bound orbits in the high-eccentricity limit and the associated flux,
	which
	contains the exact dependence on the velocity and can be used in the future to improve the
	waveform at large velocities  \cite{Buonanno:2022pgc}.
App.~\ref{app:Kin} summarizes our kinematics conventions.
App.~\ref{app:FT} details the notation for integration, Fourier transforms and index contractions.
In App.~\ref{app:PM} we quote the tidal effects in the impulse.

\paragraph*{Eikonal operator.}
The eikonal operator determines the final state of a gravitational collision in terms of the initial one in the classical limit.
It combines the eikonal phase $\operatorname{Re}2\delta$, which determines the deflection (see~\cite{DiVecchia:2021bdo} and references therein) and is sensitive to tidal effects starting at one loop \cite{Bern:2020uwk}, with the gravitational waveform $\tilde{\mathcal{A}}^{\mu\nu}$ \cite{Jakobsen:2020ksu,Mougiakakos:2020laz,Jakobsen:2021smu,Mougiakakos:2021ckm}, obtained to leading order from the five-point amplitude $\mathcal{A}^{\mu\nu}$ via Fourier transform \eqref{A}. Introducing the graviton creation/annihilation operators $\hat a_k^\dagger$, $\hat a_k$, up to 3PM order it describes gravitational waves as coherent graviton emissions, \footnote{The more precise form of the eikonal operator \cite{Cristofoli:2021jas,DiVecchia:2022piu} also involves integrations over the massive particles' phase space, but this will not be relevant for the present calculations.}
\begin{equation}\label{eikope}
	\hat{S}
	=
	e^{i\operatorname{Re}2\delta}
	e^{i\int_k\left(\tilde{\mathcal{A}}(k) \hat{a}^\dagger_k+\tilde{\mathcal{A}}^\ast(k) \hat{a}_k\right)}
\end{equation}
so that $|\Psi_\text{out}\rangle=\hat{S} |\Psi_\text{in}\rangle$,
where $|\Psi_\text{in}\rangle$ models two incoming particles with mass $m_1$, $m_2$ and impact parameter $b^\alpha=b_1^\alpha-b_2^\alpha$, while $|\Psi_\text{out}\rangle$ captures the final configuration.
In the following, we will calculate the expectation values of the linear and angular momentum operators of the gravitational field in the final state
\begin{align}\label{naiveexp}
	\boldsymbol P^\alpha
	\!=\!
	\langle \Psi_\text{in} | \hat{S}^\dagger \hat P^\alpha \hat{S} | \Psi_\text{in}\rangle,\quad
	\boldsymbol J^{\alpha\beta}
	\!=\!
	\langle \Psi_\text{in} | \hat{S}^\dagger \hat J^{\alpha\beta} \hat{S} | \Psi_\text{in}\rangle
\end{align}
taking into account tidal corrections.
Since the (connected) amplitude $\mathcal A^{\mu\nu}$ only includes the standard Weinberg limit of \emph{soft but nonzero} momenta, the quantities in \eqref{naiveexp} only include effects due to dynamically propagating gravitons, and involve no contributions localized at zero frequency, i.e.~no static terms. 

To include such terms, it is sufficient to perform the following dressing \cite{DiVecchia:2022owy,DiVecchia:2022piu},
\begin{equation}\label{}
	|\text{out/in}\rangle = e^{\int_k\left(F_\text{out/in}(k) \hat{a}^\dagger_k- F_\text{out/in}(k) \hat{a}_k\right)}|\Psi_\text{out/in}\rangle\,,
\end{equation}
where, introducing a soft scale $\omega^\ast$ (to be later sent to 0),
\begin{equation}\label{}
	F_{\text{out/in}}^{\mu\nu}(k) = \Theta(\omega^\ast-k^0) \sum_{n\in\text{out/in}} \frac{\eta_n \sqrt{8\pi G} \,p_n^\mu p_n^\nu}{p_n\cdot k-i0}\,,
\end{equation}
and $\eta_n=+1$ if $n$ is outgoing, $\eta_n=-1$ if $n$ is incoming,
which recovers the static effects via the $-i0$ prescription \cite{Mougiakakos:2021ckm,Riva:2021vnj,Manohar:2022dea,DiVecchia:2022owy}.
In this way, $	|\text{out}\rangle  = \hat{\mathcal{S}} |\text{in} \rangle$ provided
\begin{equation}\label{eikope4d}
    \hat{\mathcal{S}}
	\!=\!
	e^{i2\tilde\delta}
	e^{\int_k\left( F(k) a^\dagger_k-F^\ast(k) a_k \right)}
	e^{i\!\int_k\left( \tilde{\mathcal{A}}(k) \hat{a}^\dagger_k+\tilde{\mathcal{A}}^\ast\!(k) \hat{a}_k \right)}
\end{equation}
where $F^{\mu\nu} = F^{\mu\nu}_{\text{out}}-F^{\mu\nu}_{\text{in}}$ is the total soft factor and $2\tilde\delta=\operatorname{Re}2\delta-2\delta^\text{RR}$ is the \emph{conservative} eikonal phase \cite{DiVecchia:2022piu}.
The dressed expectation values  \cite{Mirbabayi:2016axw,Arkani-Hamed:2017jhn,Choi:2017ylo,Hannesdottir:2019opa}
\begin{align}\label{trueexp}
	P^\alpha
	\!=\!
	\langle \Psi_\text{in} | \hat{\mathcal{S}}^\dagger \hat P^\alpha \hat{\mathcal{S}} | \Psi_\text{in}\rangle,\quad
	J^{\alpha\beta}
	\!=\!
	\langle \Psi_\text{in} | \hat{\mathcal{S}}^\dagger \hat J^{\alpha\beta} \hat{\mathcal{S}} | \Psi_\text{in}\rangle
\end{align}
then also capture the effects of the static gravitational field. The distinction between \eqref{naiveexp} and \eqref{trueexp} is irrelevant for the linear momentum, $P^\alpha= \boldsymbol{P}^\alpha$, but the angular momentum is sensitive to it \cite{Damour:2020tta,Veneziano:2022zwh,DiVecchia:2022owy,DiVecchia:2022piu,Javadinezhad:2022ldc,Riva:2023xxm,Compere:2023qoa} $J^{\alpha\beta}= \boldsymbol{J}^{\alpha\beta}+\mathcal J^{\alpha\beta}$, with the former term due to radiative modes and the latter due to static modes \cite{Mougiakakos:2021ckm,Riva:2021vnj,Manohar:2022dea,DiVecchia:2022owy,DiVecchia:2022piu}.

\paragraph*{Tidal effects in the five-point amplitude.}
The $2\to3$ amplitude in the classical limit $\mathcal{A}^{\mu\nu}$ for graviton emissions up to linear order in the tidal couplings \cite{Bern:2020uwk,Cheung:2020sdj,Aoude:2020ygw} can be obtained, at tree level, from the stress-energy tensors $t^{\mu\nu}$ calculated in \cite{Mougiakakos:2021ckm,Riva:2021vnj,Mougiakakos:2022sic} via $\mathcal{A}^{\mu\nu}=4 (8\pi G)^{3/2} m_1^2 m_2^2 t^{\mu\nu} / (q_1^2 q_2^2)$ \footnote{
	To leading order, both the amplitude  $\mathcal A^{\mu\nu}$ \cite{Luna:2017dtq,DiVecchia:2021bdo,Cristofoli:2021vyo} and the stress-energy tensor obtained from worldline methods $t^{\mu\nu}$ \cite{Mougiakakos:2021ckm,Riva:2021vnj,Mougiakakos:2022sic} are directly linked to the classical waveform and, from this common quantity, one can then easily fix the relative factor. This simple connection is only valid to leading order (tree level), since appropriate subtractions are needed to obtain the subleading waveform from the one-loop amplitude \cite{Brandhuber:2023hhy,Herderschee:2023fxh,Georgoudis:2023lgf}. 
	See \eqref{Aq1q2k} for our conventions on $q_1$, $q_2$ and $k$. As anticipated, we drop static $\delta(\omega)$ contributions in $\mathcal{A}^{\mu\nu}$.}. We shall follow the notation of \cite{Mougiakakos:2022sic} and denote by $c_{E_i^2}$, $c_{B_i^2}$, where $i=1,2$ labels the two colliding objects, the couplings associated to mass/electric-type and current/magnetic-type effects, $X=E,B$ for short \footnote{See \cite{Bern:2020uwk,Cheung:2020sdj} for the mapping between field-basis and worldline-basis tidal operators.
	}.
These are related to the Love numbers $k_i^{(2)}$, $j_i^{(2)}$ by  $c_{E_i^2}=\tfrac16 k_i^{(2)}R_i^5/G$ and $c_{B_i^2}=\tfrac1{32} j_i^{(2)}R_i^5/G$  with $R_i$ the radius of object $i$. Note that $R_i=G m_i/K_i$, with $K_i$ an additional perturbative parameter characterizing the star's ``compactness'', roughly of order $0.1$--$0.2$ for typical neutron stars \cite{Blanchet:2013haa,LIGOScientific:2017vwq}. In this way, tidal effects can compete with point-particle effects,  controlled by $G m_i/b$ \footnote{With reference to Eq.~\eqref{Jcomplete}, the fraction of the initial angular momentum that is lost due to tidal effects scales like $J_\text{tid}/J\sim (Gm/b)^7 k_1^{(2)}/K_1^5$ while the fraction lost due to point-particle effects scales like $J_\text{pp}/J\sim (Gm/b)^3$ \cite{Manohar:2022dea}.}.

We first restrict for simplicity to the case $c_{X_2^2}=0$, the general case will be obtained by symmetrizing over particle labels.
Accordingly, ${\mathcal{A}}^{\mu\nu}={\mathcal{A}}_\text{pp}^{\mu\nu}+{\mathcal{A}}_{E_1^2}^{\mu\nu}+{\mathcal{A}}_{B_1^2}^{\mu\nu}$, where ${\mathcal{A}}_\text{pp}^{\mu\nu}$ is the point-particle contribution \cite{Goldberger:2016iau,Luna:2017dtq,Mogull:2020sak} and the ${\mathcal{A}}_{X_1^2}^{\mu\nu}$ capture the linear tidal effects \cite{Mougiakakos:2022sic}. We provide their expressions in an ancillary file.
Note that $\mathcal{A}^{\mu\nu}$ obeys the conservation condition only up to contact terms, $k_\mu \mathcal{A}^{\mu\nu}=C^{\nu}$, where $C^{\nu}$ is analytic in $q_1^2$ and $q_2^2$ and thus vanishes upon Fourier transform \eqref{A} for large $b$. We checked that our results are unchanged if we add contact terms to $\mathcal{A}^{\mu\nu}$.

\paragraph*{Radiative modes.}
In view of \eqref{eikope}, the formula expressing the total radiated energy-momentum \eqref{naiveexp} in terms of $\tilde{\mathcal{A}}^{\mu\nu}$ is given by \cite{Herrmann:2021lqe,Herrmann:2021tct,DiVecchia:2021bdo,Manohar:2022dea}
\begin{equation}\label{Prad}
	\boldsymbol{P}^\alpha = \int_{ k} \tilde{\mathcal{A}}\, k^\alpha \tilde{\mathcal{A}}^\ast\,.
\end{equation}
Since $\tilde{\mathcal{A}}^{\mu\nu}$ involves a complicated dependence on Bessel functions while $\mathcal{A}^{\mu\nu}$ is a rational function, it pays off to recast this integral as the Fourier transform of a convolution, i.e.~an integral over the 3-particle phase-space,
\begin{equation}\label{3pcqspacesimpl}
	\boldsymbol P^\alpha
	=\!
	\operatorname{FT}
	\!\int\!
	d(\text{LIPS})
	k^\alpha\!\!\!\!\!\!\!\!\!\!
	\begin{gathered}
		\begin{tikzpicture}[scale=.4]
			\draw[<-] (-4.8,5.17)--(-4.2,5.17);
			\draw[<-] (-1,5.15)--(-1.6,5.15);
			\draw[<-] (-1,3.15)--(-1.6,3.15);
			\draw[<-] (-1,.85)--(-1.6,.85);
			\draw[<-] (-4.8,.83)--(-4.2,.83);
			\draw[<-] (-2.85,1.7)--(-2.85,2.4);
			\draw[<-] (-2.85,4.3)--(-2.85,3.6);
			\path [draw, thick, blue] (-5,5)--(-3,5)--(-1,5);
			\path [draw, thick, color=green!60!black] (-5,1)--(-3,1)--(-1,1);
			\path [draw] (-3,3)--(-1,3);
			\path [draw] (-3,1)--(-3,5);
			\draw[dashed] (-3,3) ellipse (1.3 and 2.3);
			\node at (-1,3)[below]{$k$};
			\node at (-5,5)[left]{$p_1$};
			\node at (-5,1)[left]{$p_2$};
			\node at (-2.8,4)[left]{$q_1$};
			\draw[<-] (3.35,5.17)--(2.75,5.17);
			\draw[<-] (-.45,5.15)--(.15,5.15);
			\draw[<-] (-.45,3.15)--(.15,3.15);
			\draw[<-] (-.45,.85)--(.15,.85);
			\draw[<-] (3.35,.83)--(2.75,.83);
			\draw[<-] (1.4,1.7)--(1.4,2.4);
			\draw[<-] (1.4,4.3)--(1.4,3.6);
			\path [draw, thick, red] (-.7,.6)--(-.7,5.4);
			\path [draw, thick, blue] (3.55,5)--(1.55,5)--(-.45,5);
			\path [draw, thick, color=green!60!black] (3.55,1)--(1.55,1)--(-.45,1);
			\path [draw] (1.55,3)--(-.45,3);
			\path [draw] (1.55,1)--(1.55,5);
			\draw[dashed] (1.55,3) ellipse (1.3 and 2.3);
			\node at (1.35,4)[right]{$q-q_1$};
		\end{tikzpicture}
	\end{gathered}
\end{equation}
Here FT is the Fourier transform defined in \eqref{FT2}, each five-point amplitude represents $\mathcal{A}^{\mu\nu}$ as in \eqref{Aq1q2k} and $d(\text{LIPS})$ stands for the
Lorentz-invariant phase space measure in the soft region \cite{Parra-Martinez:2020dzs,DiVecchia:2021bdo,Herrmann:2021tct},  
\begin{equation*}\label{}
	\tfrac{d^Dk}{(2\pi)^D}2\pi\theta(k^0)\delta(k^2)
	\tfrac{d^Dq_1}{(2\pi)^D}2\pi\delta(2p_1\cdot q_1)
	2\pi\delta(2p_2\cdot (q_1 + k)).
\end{equation*}
To evaluate these integrals, we use reverse unitarity \cite{Anastasiou:2002qz,Anastasiou:2002yz,Anastasiou:2003yy,Herrmann:2021lqe,Herrmann:2021tct}. Starting from \eqref{3pcqspacesimpl}, we first rewrite the phase-space delta functions as ``cut'' propagators, via the identity $2i\pi\delta(x) = \frac{1}{x-i0}-\frac{1}{x+i0}$, and then apply Integration By Parts (IBP) identities to the resulting  integrals of rational functions to recast them as linear combinations of the Master Integrals (MIs) calculated in Refs.~\cite{DiVecchia:2020ymx,DiVecchia:2021bdo,Herrmann:2021tct}. We employ the \texttt{Mathematica} package \texttt{LiteRed} \cite{Lee:2012cn,Lee:2013mka} for the IBP reduction. We refer to \cite[Sect.~3, 6.1]{DiVecchia:2021bdo} for more details about the integration and the MIs.
Focusing on the terms linear in the tidal effects, we thus confirm the results of Refs.~\cite{Mougiakakos:2022sic,Jakobsen:2022psy},
\begin{equation}\label{Ptid}
	\boldsymbol{P}_\text{tid}^\alpha
	=
	R_f
	\sum_X \frac{c_{X_1^2}}{m_1} \left(
	\mathcal E^X \check{u}_1^\alpha
	+
	\mathcal F^X \check{u}_2^\alpha
\right),
\end{equation}
where 
$R_f = 15 \pi G^3 m_1^2 m_2^2/(64\, b^7)$, while
\begin{equation}\label{}
	\mathcal E^X = f_1^X + f_2^X \log\frac{\sigma+1}{2}+f_3^X\frac{\sigma \operatorname{arccosh}\sigma}{2\sqrt{\sigma^2-1}}\,,
\end{equation}
with $f^X_3=-\left(\sigma^2-\tfrac32\right)f_2^X/(\sigma^2-1) $, and $f_1^X$, $f_2^X$, $\mathcal F^X$
are given in Table~\ref{tab:Pradiative}
as functions of $\sigma = -u_1\cdot u_2$.

\begin{table}[b]%The best place to locate the table environment is directly after its first reference in text
	\fbox{
		\begin{minipage}{.9\linewidth}
				\vspace{-6pt}
			\begin{align*}
				f^E_1 &\!=\! \frac{(\sigma^2-1)^{-\frac{1}{2}}}{2(\sigma+1)^3}[
				937 \sigma^9\!\!+\!1551 \sigma^8\!\!\!-\!2463\sigma^7\!\!\!-\!5645 \sigma^6\!\!\\
				&+20415 \sigma^5\!\!+\!65965 \sigma^4\!\!\!-\!349541 \sigma^3\!\!+\! 535057 \sigma^2\!\!\\
				&-360356 \sigma +92160
				]\\
				f^E_2 &\!=\! 30\sqrt{\sigma^2-1} (21 \sigma ^4-14 \sigma ^2+9) \\
				\mathcal F^E &\!=\! \frac{3(\sigma^2-1)^{\frac{3}{2}}}{(\sigma+1)^5}[
				42 \sigma^8\!+\! 210 \sigma^7\!+\!315 \sigma^6\!-\!105 \sigma^5\\
				&-\!944 \sigma^4\!\!\!-\!1528 \sigma^3\!+\!22011 \sigma^2\!\!\!-\!33201 \sigma \!+\!16272
				]\\
				f^B_1 &\!=\! \frac{\sqrt{\sigma^2-1}}{4(\sigma+1)^4}[
				1559 \sigma^8\!\!+\!3716 \sigma^7\!\!\!-\! 1630 \sigma^6\!\!\!-\!11660 \sigma^5\!\!\\
				&+\!28288 \sigma^4\!\!+\!155292 \sigma^3\!\!-\!543442 \sigma^2\!+\!535212 \sigma\! -\!180775
				]\\
				f^B_2 &\!=\! 210 (\sigma^2-1)^{\frac{3}{2}} (3\sigma^2+1)\\
				\mathcal F^B &\!=\!
				 \frac{-3(
				 	105 \sigma^5\!\!+\!1630 \sigma^4\!\!+\!1840 \sigma^3\!\!+\!3690 \sigma^2\!\!\!-\!17769 \sigma\! +\!15984
				 	)}{(\sigma+1)^6(\sigma^2-1)^{-\frac{5}{2}}}
			\end{align*}
		\end{minipage}
	}
	\caption{\label{tab:Pradiative}%
		Functions entering the radiated energy-momentum due to linear tidal effects \cite{Mougiakakos:2022sic}.}
\end{table}
The expression \eqref{Ptid} for the radiated energy-momentum holds for $c_{X_2^2}=0$ and the generic case is obtained by symmetrizing over $1\leftrightarrow2$. 
The relation between $f_3^X$ and $f_2^{X}$ is due to the fact that the tidal interactions under consideration are linear (no ``H topology'') \cite{Mougiakakos:2022sic}.
Verifying that the coefficient of $b^\mu$ in \eqref{Ptid} vanishes provides an internal cross-check for the calculation. Indeed, since the integrand of \eqref{3pcqspacesimpl} is real, a component along $b^\mu$ would originate from a term of the type $f(\sigma,q^2)\,q^\mu$ with real $f$, whose Fourier transform \eqref{FT2} is purely imaginary.

From the eikonal operator \eqref{eikope}, one can derive the following formulas expressing the radiated angular momentum \eqref{naiveexp} in terms of $\tilde{\mathcal{A}}^{\mu\nu}$ \cite{Manohar:2022dea,DiVecchia:2022owy,DiVecchia:2022piu}: $\boldsymbol J_{\alpha\beta} = \boldsymbol J^{(o)}_{\alpha\beta} + \boldsymbol J^{(s)}_{\alpha\beta}$,
\begin{equation}\label{Jgravexpl}
i\boldsymbol J^{(o)}_{\alpha\beta}
=
\int_{{k}}
k_{[\alpha}
\frac{\partial\tilde{\mathcal{A}}}{\partial k^{\beta]}}\tilde{\mathcal{A}}^\ast
\,,\quad
\boldsymbol J^{(s)}_{\alpha\beta}
=
i \int_{{k}}
2\tilde{\mathcal{A}}^{\mu}_{[\alpha} \tilde{\mathcal{A}}_{\beta]\mu}^\ast\,.
\end{equation}
Under a translation \eqref{translation}, 
\cite{Manohar:2022dea}
\begin{equation}
\label{translation-J}
\boldsymbol{J}^{\alpha\beta}\to \boldsymbol J^{\alpha\beta}+a^{[\alpha} \boldsymbol{P}^{\beta]}\,.
\end{equation} 
It is straightforward to express $\boldsymbol{J}^{(s)}_{\alpha\beta}$ as the Fourier transform of a three-particle cut, as we did for $\boldsymbol{P}^\alpha$ in \eqref{3pcqspacesimpl}, with appropriate index contractions. This step is more delicate for  $\boldsymbol{J}_{\alpha\beta}^{(o)}$, which involves derivatives with respect to $k^\mu$ that can act on the mass-shell delta functions. Nevertheless, in a frame where $b_1^\mu=b^\mu$ and $b_2^\mu=0$ where \eqref{Awhenb2=0} applies, one can recast it in the form \cite{DiVecchia:2022piu}
\begin{align}
	%\begin{split}
		&i\boldsymbol{J}^{(o)}_{\alpha\beta}
		\!=\!
		\operatorname{FT}
		\!\int
		\!\! k_{[\alpha}
		\frac{\partial }{\partial k^{\beta]}}
		\!\!	\left[ d(\text{LIPS})\!\!\!\!\!\!
		\!\!\!\!\!\!
		\begin{gathered}
			\begin{tikzpicture}[scale=.4]
				\draw[<-] (-4.8,5.17)--(-4.2,5.17);
				\draw[<-] (-1,5.15)--(-1.6,5.15);
				\draw[<-] (-1,3.15)--(-1.6,3.15);
				\draw[<-] (-1,.85)--(-1.6,.85);
				\draw[<-] (-4.8,.83)--(-4.2,.83);
				\draw[<-] (-2.85,1.7)--(-2.85,2.4);
				\draw[<-] (-2.85,4.3)--(-2.85,3.6);
				\path [draw, thick, blue] (-5,5)--(-3,5)--(-1,5);
				\path [draw, thick, color=green!60!black] (-5,1)--(-3,1)--(-1,1);
				\path [draw] (-3,3)--(-1,3);
				\path [draw] (-3,1)--(-3,5);
				\draw[dashed] (-3,3) ellipse (1.3 and 2.3);
				\node at (-1,3)[below]{$k$};
				\node at (-5,5)[left]{$p_1$};
				\node at (-5,1)[left]{$p_2$};
				\node at (-2.8,4)[left]{$q_1$};
			\end{tikzpicture}
		\end{gathered}
	\!\!
		\right]
		\!\!\!
		\begin{gathered}
			\begin{tikzpicture}[scale=.4]
				\draw[<-] (4.8,5.17)--(4.2,5.17);
				\draw[<-] (1,5.15)--(1.6,5.15);
				\draw[<-] (1,3.15)--(1.6,3.15);
				\draw[<-] (1,.85)--(1.6,.85);
				\draw[<-] (4.8,.83)--(4.2,.83);
				\draw[<-] (2.85,1.7)--(2.85,2.4);
				\draw[<-] (2.85,4.3)--(2.85,3.6);
				\path [draw, thick, blue] (5,5)--(3,5)--(1,5);
				\path [draw, thick, color=green!60!black] (5,1)--(3,1)--(1,1);
				\path [draw] (3,3)--(1,3);
				\path [draw] (3,1)--(3,5);
				\draw[dashed] (3,3) ellipse (1.3 and 2.3);
				\node at (2.8,4)[right]{$q-q_1$};
			\end{tikzpicture}
		\end{gathered}
		\nonumber
		\\
		\label{LL1Lp}
		&
		-\!
		u_{2[\alpha}
		\!
		\operatorname{FT}\! 
		\frac{\partial}{\partial q_{\parallel2}}
		\!\int\!\!  d(\text{LIPS})
		k^{\phantom{(2)}}_{\beta]}\!\!\!\!\!\!\!\!\!\!\!\!\!\!\!\!\!
		\begin{gathered}
		\begin{tikzpicture}[scale=.4]
			\draw[<-] (-4.8,5.17)--(-4.2,5.17);
			\draw[<-] (-1,5.15)--(-1.6,5.15);
			\draw[<-] (-1,3.15)--(-1.6,3.15);
			\draw[<-] (-1,.85)--(-1.6,.85);
			\draw[<-] (-4.8,.83)--(-4.2,.83);
			\draw[<-] (-2.85,1.7)--(-2.85,2.4);
			\draw[<-] (-2.85,4.3)--(-2.85,3.6);
			\path [draw, thick, blue] (-5,5)--(-3,5)--(-1,5);
			\path [draw, thick, color=green!60!black] (-5,1)--(-3,1)--(-1,1);
			\path [draw] (-3,3)--(-1,3);
			\path [draw] (-3,1)--(-3,5);
			\draw[dashed] (-3,3) ellipse (1.3 and 2.3);
			\node at (-1,3)[below]{$k$};
			\node at (-5,5)[left]{$p_1$};
			\node at (-5,1)[left]{$p_2$};
			\node at (-2.8,4)[left]{$q_1$};
			\draw[<-] (3.35,5.17)--(2.75,5.17);
			\draw[<-] (-.45,5.15)--(.15,5.15);
			\draw[<-] (-.45,3.15)--(.15,3.15);
			\draw[<-] (-.45,.85)--(.15,.85);
			\draw[<-] (3.35,.83)--(2.75,.83);
			\draw[<-] (1.4,1.7)--(1.4,2.4);
			\draw[<-] (1.4,4.3)--(1.4,3.6);
			\path [draw, thick, red] (-.7,.6)--(-.7,5.4);
			\path [draw, thick, blue] (3.55,5)--(1.55,5)--(-.45,5);
			\path [draw, thick, color=green!60!black] (3.55,1)--(1.55,1)--(-.45,1);
			\path [draw] (1.55,3)--(-.45,3);
			\path [draw] (1.55,1)--(1.55,5);
			\draw[dashed] (1.55,3) ellipse (1.3 and 2.3);
			\node at (1.35,4)[right]{$q-q_1$};
		\end{tikzpicture}
	\end{gathered}
	%\end{split}
\end{align}
where the derivative in the first line can act both on $\mathcal A^{\mu\nu}$ and on $d(\text{LIPS})$, 
and $q_{\parallel2}=-u_2\cdot q^\mu$. The integrals to be performed then belong to the same family as for \eqref{3pcqspacesimpl}, so we can evaluate them in the same way.

Translating the result to a frame where $b_1^\mu=0$ and $b_2^\mu=-b^\mu$, using the simple transformation law \eqref{translation-J} and the explicit result \eqref{Ptid} for $\boldsymbol{P}_\text{tid}^\alpha$, we obtain the following new result for the radiated angular momentum due to linear tidal effects,
\begin{equation}\label{Jtid}
	\boldsymbol{J}_\text{tid}^{\alpha\beta}
	=
R_f \sum_X
\frac{c_{X_1^2}}{m_1}
	\left(
	\mathcal C^X b^{[\alpha}u_1^{\beta]}
	+
	\mathcal D^X u_2^{[\alpha} b^{\beta]}
	\right)
\end{equation}
where $R_f$ is given below Eq.~\eqref{Ptid},
\begin{align}
	\mathcal C^X &\!=\! g^X_1 + g^X_2 \log\frac{\sigma+1}{2} + g^X_3 \frac{\sigma\,\operatorname{arccosh}\sigma}{2\sqrt{\sigma^2-1}}\,, \\
	\mathcal D^X &\!=\! h^X_1 + h^X_2 \log\frac{\sigma+1}{2} + h^X_3 \frac{\sigma\,\operatorname{arccosh}\sigma}{2\sqrt{\sigma^2-1}}\,,
\end{align}
with $g^X_3\!=\!-\left(\sigma^2-\tfrac32\right)g_2^X/(\sigma^2-1)$ and similarly $h^X_3\!=\!-\left(\sigma^2-\tfrac32\right)h_2^X/(\sigma^2-1)$,
and the functions $g_j^X$, $h_j^X$ for $j=1,2$ are detailed in Table~\ref{tab:Jradiative} for $X=E$ and in Table~\ref{tab:JradiativeB} for $X=B$. 
\begin{table}[b]%The best place to locate the table environment is directly after its first reference in text
	\fbox{
		\begin{minipage}{.9\linewidth}
			\vspace{-5pt}
			\begin{align*}
				g^E_1 &\!=\! \frac{(\sigma^2-1)^{-\frac{3}{2}}}{10(\sigma+1)^3}
				\big[
				2573 \sigma^9\!\!+\!9819 \sigma^8\!\!+\!13143 \sigma^7\!\!
				+\!1845 \sigma^6\\
				&-\!897603 \sigma^5\!\!+\!3221239 \sigma^4\!\!\!-\!5046195 \sigma^3\!\!+\!4203751 \sigma^2\\
				&-\!1862318 \sigma +351826
				\big]\\
				g^E_2 &\!=\! -{6(35 \sigma^4-50 \sigma^2-1)}/{\sqrt{\sigma^2-1}} \\
				h^E_1 &\!=\! \frac{4(\sigma^2-1)^{-\frac{3}{2}}}{5(\sigma+1)^2}
				[
				492 \sigma^7\!\!+\!564 \sigma^6\!\!\!-\!609 \sigma^5\!\!\!-\!722 \sigma^4\\
				&-\!4636 \sigma^3\!+\!13478 \sigma^2\!-\!14143 \sigma \!+\!5096
				]\\
				h^E_2 &\!=\! {48\sigma(7\sigma^2+1)}/{\sqrt{\sigma^2-1}} 
			\end{align*}
		\end{minipage}
	}
	\caption{\label{tab:Jradiative}%
		Functions entering the radiated angular momentum due to $E_1^2$ tidal coupling.}
\end{table}
\begin{table}[b]%The best place to locate the table environment is directly after its first reference in text
	\fbox{
		\begin{minipage}{.9\linewidth}
			\vspace{-5pt}
			\begin{align*}
				g^B_1 &\!=\! \frac{20(\sigma^2-1)^{-\frac{1}{2}}}{(\sigma+1)^4}[
			4495 \sigma^8\!\!+\!22180 \sigma^7\!\!+\!46630 \sigma^6\!\!+\!50020 \sigma^5\\
			-&1748636 \sigma^4\!\!+\!4687932 \sigma^3\!\!\!-\!5397990 \sigma^2\!\!+\!3026428 \sigma \!-\!681459
			]\\
			g^B_2 &\!=\! - 30 \sqrt{\sigma^2-1}  (7 \sigma ^2-3)\\
			h^B_1 &\!=\! \frac{2(\sigma^2-1)^{-\frac{1}{2}}}{5(\sigma+1)^3}
			[
			879 \sigma^6\!\!+\!1797 \sigma^5\!\!\!-\!492 \sigma^4\!\!\!-2908 \sigma^3\\
			&-\!10491 \sigma^2\!\!+\!18815 \sigma \!-\! 9280
			]\\
			h^B_2 &\!=\! 336 \sigma \sqrt{\sigma^2-1} 
			\end{align*}
		\end{minipage}
	}
	\caption{\label{tab:JradiativeB}%
		Functions entering the radiated angular momentum due to $B_1^2$ tidal coupling.}
\end{table}
Verifying that the coefficient of $u_1^{[\alpha}u_2^{\beta]}$ in \eqref{Jtid} vanishes serves an internal consistency check analogous to one discussed for $\boldsymbol{P}_\text{tid}^\alpha$. Each integrand on the right-hand side of \eqref{LL1Lp} is real and terms $f(\sigma,q^2)\,u_{1}^{[\alpha} u_2^{\beta]}$ with real $f$ would be real in $b$-space, and hence contribute imaginary terms to the radiated angular momentum. In fact, such terms do appear separately in each line of \eqref{LL1Lp}, but they crucially cancel out in the sum. 

Since \eqref{Jtid} holds for $c_{X_2^2}=0$ in a frame where $b_1^\alpha=0$, $b_2^\alpha=-b^\alpha$, interchanging all particle labels in it yields the radiated angular momentum for $c_{X_1^2}= 0$ in a frame where $b_1^\alpha=b^\alpha$, $b_2^\alpha=0$ instead, but one can obtain $\boldsymbol{J}_\text{tid}^{\alpha\beta}$ in any desired translation frame with the help of the simple transformation law \eqref{translation-J} and the explicit form \eqref{Ptid} for $\boldsymbol{P}^\alpha_\text{tid}$.

In addition to translations, Eq.~\eqref{Jtid} is also covariant under Lorentz transformations.
The physical meaning of $\mathcal C^X$ and $\mathcal D^X$ becomes transparent in frames where not only $b_1^\alpha=0$ but one of the two particles is also initially at rest, where they are proportional to the angular momentum of gravitational waves. For definiteness, we align the impact parameter along the $y$ axis, $b^\alpha=(0,0,b,0)$, and the motion of the incoming particle along the $x$ axis. In a frame where particle 1 is at rest, $u_1^\alpha=(1,0)$, $u_2^\alpha=(\sigma,p_\infty,0,0)$ with $p_\infty= \sqrt{\sigma^2-1}$, so
\begin{equation}\label{}
	\boldsymbol{J}_\text{tid}^{xy} = R_f b p_\infty\,\sum_X \frac{c_{X_1^2}}{m_1}\,\mathcal D^X\,,
\end{equation} 
while in a frame where particle 2 is at rest $u_1^\alpha=(\sigma,-p_\infty,0,0)$, $u_2^\alpha=(1,0)$, the same formula applies with $\mathcal{C}^X$ replaced by $\mathcal D^X$.
In the nonrelativistic limit $p_\infty\to0$,
\begin{equation}\label{CDPN}
	\begin{split}
	\mathcal{C}^E &= \tfrac{1056}{5}p_{\infty}-\tfrac{349}{35} p_{\infty}^3+\mathcal O(p_\infty^5)\,\\
	\mathcal{D}^E &= \tfrac{1056}{5}p_{\infty}-\tfrac{324}{7} p_{\infty}^3 +\mathcal O(p_{\infty}^5)\,\\
	\mathcal{C}^B &= 40 p_{\infty}^3+\tfrac{3833}{35}p_{\infty }^5  +\mathcal O(p_{\infty}^7)\,\\
	\mathcal{D}^B &= -\tfrac{168 }{5}p_{\infty}^3+\tfrac{1471 }{10}p_{\infty }^5 +\mathcal O(p_\infty^7) \,.
	\end{split}
\end{equation}
As expected, in this limit, $B$ contributions are suppressed by an extra power of $p_\infty^2\sim v^2$ compared to $E$-type ones.

Let us now start again from Eq.~\eqref{Jtid}, which holds in a frame where $b^\alpha_1=0$, and perform a translation $b^\mu_i\to b'^\mu_i= b^\mu_i+a^\mu$ that places the center of mass (or ``center of energy'') in the origin of the transverse plane,
$(1-w) b'^\mu_1+w b'^\mu_2=0$ with $w=p_2\cdot(p_1+p_2)/(p_1+p_2)^2$.
This sets $a^\mu= wb^\mu$ and by the transformation law \eqref{translation-J} we can find 
the radiated angular momentum tensor in this new frame, $\boldsymbol{J}_\text{tid}'^{\alpha\beta}$. Its expression is obtained from \eqref{Jtid} by replacing $\mathcal{C}^X\to \mathcal{C}'^{X}=\mathcal{C}^X+w \check{\mathcal{E}}^X$ and $\mathcal{D}^X\to\mathcal{D}'^X=\mathcal D^X-w\check{\mathcal{F}}^X$ 
where $\mathcal E^X = \sigma\check{\mathcal{F}}^X + \check{\mathcal{E}}^X$ and $\mathcal F^X = \sigma\check{\mathcal{E}}^X + \check{\mathcal{F}}^X$.
Moreover, going to a frame where the center of mass is also at rest, say $b=(0,0,b,0)$, $-p_1=(E_1,-p,0,0)$, $-p_2=(E_2,p,0,0)$,
we find, for the component of the angular momentum orthogonal to the scattering plane,
\begin{equation}\label{JtidCM}
		 \boldsymbol{J}_\text{tid}\equiv
	\boldsymbol{J}_\text{tid}'^{xy}
	=
	R_f J
	\sum_X
	\frac{c_{X_1^2}}{m_1}\left(
		\frac{\mathcal{C}'^X}{m_1}
		+
		\frac{\mathcal{D}'^X }{m_2}
		\right)
		\,,
\end{equation}
where $J = pb$ is the initial angular momentum in the center-of-mass frame.
At low energies, for small $p_\infty$, we find, introducing the symmetric mass ratio $\nu= m_1m_2/m^2$, with $m=m_1+m_2$,
and  $\Delta=(m_1-m_2)/m$,
\begin{align}\label{}
	\nonumber
		\frac{\boldsymbol{J}_\text{tid}}{R_fbp_\infty} 
		&\!=\!\frac{c_{E_1^2}}{m_1}\!\left[\frac{1056}{5}  p_{\infty }\!+\!\left(\!\frac{6}{5} (49-328 \nu)\!-\!\frac{3678 \Delta }{35}\!\right)\! p_{\infty }^3\!\right]\\
		&+\frac{c_{B_1^2}}{m_1}\left(\frac{96 \Delta }{5}-\frac{264}{5}\right) p_{\infty }^3+\mathcal O(p_\infty^5)\,.
\end{align}
In the formal ultrarelativistic limit $\sigma\to\infty$ instead 
\begin{equation}\label{}
	\begin{split}
	\frac{\boldsymbol{J}_\text{tid}}{R_fJ} 
	\!=\!\frac{c_{E_1^2}}{m_1^2}\,63 \sigma^5\!\!\!-\!\frac{c_{E_1^2}+c_{B_1^2}}{2 m_1 m_2}\,315 \sigma^4\! \log \sigma
	\!+\!\mathcal O(\sigma^4).
	\end{split}
\end{equation}
Taking into account that the leading deflection angle scales as  $\Theta_s\sim G m\sqrt{\sigma}/b$, and that $c_{E_1^2}\sim G^4 m^5$, we see that $\boldsymbol{J}_\text{tid}/J\sim \Theta_s^4 (\sqrt{\sigma} \Theta_s)^3$. Therefore, if we were to take $\sigma$ arbitrarily large for fixed small $\Theta_s$, the system could radiate an arbitrarily large amount of angular momentum.
The perturbative PM expansion is however limited to $\sqrt{\sigma}\Theta_s\lesssim 1$ \cite{DEath:1976bbo,Kovacs:1977uw,Kovacs:1978eu,DiVecchia:2022nna}, so that the true ultrarelativistic limit lies beyond the scope of these calculations.

\paragraph*{Consistency check.}
To obtain a cross-check of the functions in Tables~\ref{tab:Jradiative}, \ref{tab:JradiativeB}, we perform an independent calculation of the angular momentum in the Post-Newtonian (PN) limit as in Ref.~\cite{Manohar:2022dea}. We start from $\mathcal{A}^{\mu\nu}(k)$ and expand it for small $p_\infty$ in the relevant scaling region $k^\alpha \sim\mathcal{O}(p_\infty)$ \cite{Kovacs:1977uw,Kovacs:1978eu,Mougiakakos:2021ckm}. We then perform the Fourier transform \eqref{A} term by term in the PN expansion in the frame $b_1^\alpha=0$. Finally, we substitute into the expression for $\boldsymbol{J}^{\alpha\beta}$ \eqref{Jgravexpl} and directly perform the integration over $k$, without using reverse unitarity. This involves integrals of Bessel functions, conveniently evaluated in \texttt{Mathematica}. Contracting \eqref{Jtid} with $(u_{2\alpha}-u_{1\alpha})b_\beta$  and $u_{1\alpha} b_\beta$, we obtain the small-$p_\infty$ expansions  $\tfrac12(\mathcal{C}^{E}+\mathcal{D}^E) = \frac{1056}{5}\,p_\infty + o(p_\infty)$,
$\mathcal{C}^{E}-\mathcal{D}^E = o(p_\infty)$,
and
$\tfrac12(\mathcal{C}^{B}+\mathcal{D}^B) = \frac{16}{5} p_\infty^3 + o(p_\infty^3)$,
$\mathcal{C}^{E}-\mathcal{D}^E = \frac{368}{5}\,p_\infty^3 + o(p_\infty^3)$, 
in perfect agreement with \eqref{CDPN}.
To obtain these results it is enough to retain the leading PN waveform for $E$ contributions $\mathcal{O}(p_\infty)$, while it is necessary to resolve also the first subleading correction for point-particle $\mathcal{O}( p_\infty^{-1}) + \mathcal{O}(p_\infty^0$) and $B$ contributions $\mathcal{O}(p_\infty^2)+\mathcal{O}(p_\infty^3)$. 

\paragraph*{Static modes.}
We now complete the result for the angular momentum loss by adding the zero-frequency contribution, i.e.~the effect of the static gravitational field, which arises when calculating the expectation value on dressed states \eqref{trueexp} using the eikonal operator \eqref{eikope4d},
\begin{equation}\label{calJab}
	\mathcal J_{\alpha\beta} = -i\int_{ k}\left( F^\ast\,k_{[\alpha}\frac{\partial F}{\partial k^{\beta]}}+2F^{\ast}_{\mu[\alpha}F^\mu_{\beta]}\right).
\end{equation}
To this end, we use Eq.~(3.30) of \cite{DiVecchia:2022owy}, which provides this contribution for a generic gravitational process.  Indeed, Eq.~\eqref{calJab} relies only on the form of the leading soft factor, which is universal, and thus the resulting  expression holds independently of the details of the collision.
In terms of the coefficients defined in Table~\ref{tab:omega=0},
\begin{table}[b]
	\fbox{
		\begin{minipage}{.9\linewidth}
			\vspace{-5pt}
			\begin{align*}
				\sigma_{nm}&=-\eta_n\eta_m\,\frac{p_n\cdot p_m}{m_n m_m}\,,\\
				\Delta_{nm}&=\frac{\operatorname{arccosh}\sigma_{nm}}{\sqrt{\sigma^2_{nm}-1}}\,,\\
				c_{nm} &= 
				2G\left[ \left(
				\sigma_{nm}^2-\tfrac12
				\right) \frac{\sigma_{nm}\Delta_{nm}-1}{\sigma_{nm}^2-1}
				-
				2
				\sigma_{nm}\Delta_{nm}
				\right],\\
				2\mathcal G
				&=
				c_{14}
				+
				c_{23}
				-
				2c_{13}
				\\
				\tfrac12\,\mathcal{I}&=\frac{8-5\sigma^2}{3(\sigma^2-1)}
				+
				\frac{\sigma(2\sigma^2-3)\operatorname{arccosh}\sigma}{(\sigma^2-1)^{3/2}}\,.
			\end{align*}
		\end{minipage}
	}
	\caption{\label{tab:omega=0}%
		Functions and coefficients entering the static terms. Here $n,m=1,2,3,4$ and $\eta_n=+1$ ($\eta_n=-1$)  if the $n$th state is outgoing (incoming).}
\end{table} 
the angular momentum due to static modes then evaluates to \cite{DiVecchia:2022owy}
\begin{equation}\label{Jabnm}
	\mathcal J^{\alpha\beta} = - \sum_{n=1,2}\sum_{m=3,4} c_{nm}\,p_n^{[\alpha} p_m^{\beta]}\,.
\end{equation}
Like the leading soft theorem, this result only depends on the momenta of the hard particles, and to obtain explicit expressions it is sufficient to substitute $p_4^\alpha = Q^\alpha - p_1^\alpha$, $p_3=-Q^\alpha-p_2^\alpha$ and the PM expansion of the impulse $Q$.

Let us note that the static contribution \eqref{Jabnm} is invariant under translations and covariant under Lorentz transformations.
In the center-of-mass frame (aligning the axes as above), we find \cite{DiVecchia:2022owy}
$\mathcal{J}^{xy} = G pQ\, \mathcal I$ up to $\mathcal O(G^4)$ corrections,
with $\mathcal I$ given in Table~\ref{tab:omega=0}.
In view of the overall power of $G$, since the $\mathcal O(G)$ impulse is unaffected by tidal terms, there is no tidal angular momentum loss to $\mathcal O(G^2 c_{X_1^2})$, and therefore (via linear response \cite{Bini:2018ywr,Bini:2021gat}) no tidal radiation-reaction in the deflection angle to $\mathcal O(G^3 c_{X_1^2})$, as noted in \cite{Mougiakakos:2022sic}.
The leading tidal effects in $\mathcal J^{xy}$ are $\mathcal O(G^3 c_{X_1^2})$ and can be obtained by substituting \eqref{PM} in it, finding the following new result
\begin{equation}\label{jstatCM}
	\mathcal{J}_\text{tid} = R_f J (\hat Q_{E_1^2}+\hat Q_{B_1^2}) \mathcal I 
\end{equation}
up to $\mathcal O(G^4 c_{X_1^2})$ corrections.
The leading, i.e.~$\mathcal O(G^4)$, tidal radiation reaction on the angle or impulse due to static modes, $\mathcal{Q}^\alpha$, can be obtained by \cite{DiVecchia:2022piu} $\mathcal Q^{\alpha}
= \frac12\frac{\partial Q^2}{\partial b_\alpha} 
\mathcal G$.
To leading order in the tidal effects \eqref{PM}, we then have $\mathcal Q^\alpha_\text{tid}= -b^\alpha \mathcal{Q}_\text{tid}/{b}$ with
$\mathcal Q_\text{tid}
= 
\tfrac{7}{2}G b^{-1}\,Q_\text{1PM}(Q_{E_1^2}+Q_{B_1^2})\,\mathcal I$.
This result agrees with the one obtained by applying the linear-response formula \cite{Bini:2012ji,Bini:2021gat}
$\mathcal Q_\text{tid} = -\frac{1}{2}\frac{\partial Q}{\partial J}\,\mathcal{J}_\text{tid}$.
Expanding for small $p_\infty$ one finds
\begin{equation}\label{}
	\begin{split}
	\frac{m}{2m_2}
	\frac{\mathcal{J}_\text{tid}}{R_fbp_\infty} 
	&\!=\!\frac{c_{E_1^2}}{m_1}\!\left[\frac{384}{5} p_{\infty}-\frac{192}{35} (7 \nu -29) p_{\infty }^3\right]\\
	&+\frac{c_{B_1^2}}{m_1}192p_\infty^3+\mathcal O(p_\infty^5)
	\end{split}
\end{equation}
while as $\sigma\to\infty$
\begin{equation}\label{}
	\begin{split}
		\frac{\mathcal{J}_\text{tid}}{R_f J} 
		\!\approx\!\frac{c_{E_1^2}+c_{B_1^2}}{m_1^2}\!\left[
		420 \sigma^3 \log\sigma
		+
		70 \sigma^3\! (6\log2\!-\!5)
		\right]
	\end{split}
\end{equation}
up to $\mathcal{O}(\sigma\log\sigma)$ corrections. 

\paragraph*{Complete result and analytic continuation.}
The total angular momentum of the gravitational field due to tidal effects constitutes the main original result of the paper and is given by the sum of the radiative piece \eqref{Jtid} and the static piece \eqref{Jabnm}. In the center-of-mass frame, it reads $J_{\text{tid}} = \boldsymbol{J}_{\text{tid}} + \mathcal{J}_{\text{tid}}$ (see Eqs.~\eqref{JtidCM}, \eqref{jstatCM}), so that
\begin{equation}\label{Jcomplete}
	J_\text{tid}= R_f J A(\sigma)\,,
\end{equation}
where we isolated a $b$-independent function $A(\sigma)$,
\begin{equation}\label{}
	A(\sigma) = 
	\sum_X
	\frac{c_{X_1^2}}{m_1}\left(
	\frac{\mathcal{C}'^X}{m_1}
	+
	\frac{\mathcal{D}'^X }{m_2}
	\right)
	+
	(\hat Q_{E_1^2}+\hat Q_{B_1^2}) \mathcal I\,.
\end{equation}

Following \cite{Kalin:2019rwq,Kalin:2019inp,Herrmann:2021tct,Saketh:2021sri,Cho:2021arx,Mougiakakos:2022sic}, one can also use \eqref{Jcomplete} to obtain the angular momentum that is radiated by a bound system in the high eccentricity limit (large $J$) during one orbital revolution. The first step is to write the total momentum radiated in the center-of-mass frame as $J_{\text{tid}}(J,\sigma) = J^{-6} f(\sigma)$ where
\begin{equation}\label{}
	f(\sigma) = \frac{15\pi G^3 m^{11}\nu^9}{64 h^{7}}\,\left(\sigma^2-1\right)^{7/2}A(\sigma)
\end{equation} 
with $h= \sqrt{1+2\nu(\sigma-1)}$.
The function $f(\sigma)$ is analytic for $\operatorname{Re}\sigma>-1$, as one can easily check since it only involves rational combinations, together with the functions $\log\frac{\sigma+1}{2}$ and 
$
	\frac{\operatorname{arccosh}\sigma}{\sqrt{\sigma^2-1}}=\frac{\operatorname{arccos}\sigma}{\sqrt{1-\sigma^2}}
$ that only have branch cuts for $\sigma<-1$.
Using the boundary-to-bound map
$J^\text{bound}_{\text{tid}}(J,\sigma) = J_{\text{tid}}(J,\sigma) + J_{\text{tid}}(-J,\sigma)$ \cite{Saketh:2021sri,Cho:2021arx},
we find
\begin{equation}\label{}
		J^\text{bound}_{\text{tid}}(J,\sigma) = \frac{2}{J^6}	\tilde f(\sigma)\,,
\end{equation}
with $\tilde{f}(\sigma)$ the analytic continuation of $f(\sigma)$ to the interval $-1<\sigma<1$. In this fashion, $J^\text{bound}_\text{tid}(J,\sigma)$ gives the leading tidal correction, for large $J$, to the angular momentum loss for bound orbits with energy $E=mh<m$.

\paragraph*{Angular momentum flux.}
Let us assume that the relation between the tidal angular momentum loss and the averaged flux $F_\text{tid}$ in isotropic gauge reads \cite{Peters:1964zz,Thorne:1980ru, Damour:1981bh,Bonga:2018gzr,Blanchet:2018yqa,Cho:2021arx,Mougiakakos:2022sic,AbhishekChowdhuri:2022ora}
\begin{equation}\label{JfluxJ}
	J_\text{tid}
	=
	\int_{-\infty}^{+\infty} \! F_\text{tid}(r,\sigma)dt
	=
	2
	\int_b^\infty \!F_\text{tid}(r,\sigma)\frac{dr}{\dot r}\,,
\end{equation}
where to leading order we can employ the straight-line trajectory, $r^2(t)\simeq b^2+v_\text{rel}^2t^2$. Here $v_\text{rel}=p/(\xi E)$ with $\xi=E_1E_2/(E_1+E_2)^2$.
Dimensional analysis fixes the $r$-dependence of the flux to be $F_\text{tid}\sim 1/r^7$. Performing the integral \eqref{JfluxJ} by noting that $2\int_b^\infty dr/(r^7\dot r)=16/(15 v_\text{rel} b^6)$ \footnote{This can be obtained by first noting that $\dot r = v^2_\text{rel} t/r= v_\text{rel}\sqrt{1-(b/r)^2}$, and then performing the change of variable $r = b/\sin\theta$.} and matching to \eqref{Jcomplete}  then determines the overall $r$-independent factor and yields 
\begin{equation}\label{}
	F_\text{tid}= \frac{225 \pi G^3   m^5 \nu ^4 \left(\sigma ^2-1\right)}{1024 h^3 \xi  r^7}\, A(\sigma)\,.
\end{equation}

\paragraph*{Conclusions.}
In this paper, we obtained a new result for the total angular momentum that is lost during a two-body scattering due to linear tidal effects, exploiting amplitude-based methods. We also provided the corresponding flux and the analytic continuation to bound orbits.
This work opens up several avenues for future work. 
A natural generalization concerns the dissipation of angular momentum in scattering with spin \cite{Riva:2022fru} and in supersymmetric theories \cite{DiVecchia:2021bdo,Herrmann:2021tct}. 
For bounded binaries, it would also be interesting to further compare with the PN literature \cite{Henry:2019xhg,Henry:2020ski,Henry:2020pzq} by performing a suitable eccentricity resummation needed to access the regime of quasicircular orbits \cite{Kalin:2019rwq,Kalin:2019rwq,Cho:2021arx}. 
A crucial next step will be to study quantitatively the impact of the present results on waveform models \cite{Antonelli:2019ytb,Khalil:2022ylj} and, of course, to extend them  by calculating $J^{\alpha\beta}$ to subleading order, three loops on the amplitude side.

\begin{acknowledgments}
I would like to thank Paolo Di Vecchia, Kays Haddad, Massimiliano M. Riva and Rodolfo Russo for very useful discussions and comments on this draft.
This work is supported by the Knut and Alice Wallenberg Foundation under grant KAW 2018.0116. Nordita is partially supported by Nordforsk.
\end{acknowledgments}

\appendix

\section{Kinematics}
\label{app:Kin}
All momenta are regarded as outgoing, so  $-p_i$ for $i=1,2$ are the physical momenta of the incoming states. The Minkowski metric is $\eta_{\mu\nu}=\text{diag}(-,+,+,+)$, so that $p_i^2+m_i^2=0$.
Four-velocities are defined by $u_{i}^\mu=-p_i^\mu/m_i$, with $u_i^2=-1$. We denote the relative Lorentz factor by $\sigma = -u_1\cdot u_2=1/\sqrt{1-v^2}$, and $v$ is the speed of body 1 as seen from the rest frame of body 2 (or vice-versa). A useful variable in the PN limit is $p_\infty= \sqrt{\sigma^2-1}$.  The spatial momentum in the center-of-mass frame is instead denoted by $p$.
It is also convenient to define variables $\check{u}_i^\mu$ which obey $\check u_i\cdot u_j =-\delta_{ij}$ by letting 
$u_1^\mu = \sigma \check u_2^\mu+ \check u_1^\mu$ and
$u_2^\mu = \sigma \check u_1^\mu+ \check u_2^\mu$.
The relative impact parameter is defined by $b^\mu= b_1^\mu-b_2^\mu$, where $b_1^\mu$ and $b_2^\mu$ are the impact parameters of each particle, transverse to the incoming directions $b_i\cdot u_j=0$.
Finally, the symmetric mass ratio is defined by $\nu= m_1m_2/m^2$, with $m=m_1+m_2$,
	while we let $\Delta=(m_1-m_2)/m$.

\section{Integration and Index Contraction}
\label{app:FT}
We employ the shorthand notation
\begin{align}\label{kLIPS}
	\int_{k} &= \int \frac{d^Dk}{(2\pi)^D}\,2\pi\theta(k^0)\delta(k^2)\,.
\end{align}
Moreover, we define
\begin{equation}\label{Aa}
	\tilde{\mathcal{A}}(k)\,\hat a^\dagger_k=\sum_i \epsilon^{(i)}_{\mu\nu}(k)^\ast\tilde{\mathcal{A}}^{\mu\nu}(k)\, \hat{a}^\dagger_i(k)
\end{equation}
where $i=1,2$ labels the two physical graviton polarizations, with polarization ``tensors'' $\epsilon^{(i)}_{\mu\nu}(k)$, and similarly for the Hermitian conjugate of \eqref{Aa} and for analogous expressions involving $F^{\mu\nu}$. The creation and annihilation operators obey canonical commutation relations,
\begin{equation}\label{}
	2\pi\theta(k^0)\delta(k^2)[\hat a_i(k),\hat a^\dagger_j(k')] \!=\! (2\pi)^{\!D}\delta^{(D)}(k-k')\delta_{ij}\,.
\end{equation}
For convenience, we suppress contractions between five-point amplitudes (unless written otherwise), letting 
\begin{equation}\label{suppressedindices}
	\mathcal A\, \mathcal A' = \mathcal A_{\mu\nu}\, \mathcal A'^{\mu\nu}-\frac1{D-2}\,\mathcal A^{\mu}_\mu \mathcal A'^\nu_\nu,
\end{equation}
and similarly for $F^{\mu\nu}$.

We define the Fourier transform $\operatorname{FT} \mathcal M$ by
\begin{equation}\label{FT2}
\operatorname{FT}\mathcal{M}\!= \!\!\int \!\!\!\tfrac{d^Dq}{(2\pi)^D}\,2\pi\delta(2p_1\cdot q)2\pi\delta(2p_2\cdot q)\,e^{ib\cdot q}\mathcal M(q)\,.
\end{equation}
The relation between the momentum-space $2\to3$ amplitude in the classical limit (the drawing inside the dashed bubble only serves as a visual help to recall the definition of $q_1$, $q_2$ and does not represent an actual Feynman diagram)
\begin{equation}\label{Aq1q2k}
	\mathcal{A}^{\mu\nu} (q_1,q_2,k)= 	\begin{gathered}
		\begin{tikzpicture}[scale=.4]
			\draw[<-] (-4.8,5.17)--(-4.2,5.17);
			\draw[<-] (-1,5.15)--(-1.6,5.15);
			\draw[<-] (-1,3.15)--(-1.6,3.15);
			\draw[<-] (-1,.85)--(-1.6,.85);
			\draw[<-] (-4.8,.83)--(-4.2,.83);
			\draw[<-] (-2.85,1.7)--(-2.85,2.4);
			\draw[<-] (-2.85,4.3)--(-2.85,3.6);
			\path [draw, thick, blue] (-5,5)--(-3,5)--(-1,5);
			\path [draw, thick, color=green!60!black] (-5,1)--(-3,1)--(-1,1);
			\path [draw] (-3,3)--(-1,3);
			\path [draw] (-3,1)--(-3,5);
			\draw[dashed] (-3,3) ellipse (1.3 and 2.3);
			\node at (-1,3)[below]{$k$};
			\node at (-5,5)[left]{$p_1$};
			\node at (-5,1)[left]{$p_2$};
			\node at (-2.8,4)[left]{$q_1$};
			\node at (-2.8,2)[left]{$q_2$};
		\end{tikzpicture}
	\end{gathered}
\end{equation}
and its $b$-space counterpart $\tilde{\mathcal{A}}^{\mu\nu}(k)$ is given by
\begin{equation}\label{A}
	\begin{split}
\tilde{\mathcal{A}}^{\mu\nu}(k) &= \!\!\int\!\!\! \frac{d^Dq_1}{(2\pi)^D}\,2\pi\delta(2p_1\cdot q_1)2\pi\delta(2p_2\cdot q_2)\\
&\times e^{ib_1\cdot q_1+ib_2\cdot q_2}\mathcal A^{\mu\nu}(q_1,q_2,k)\,,
	\end{split}
\end{equation}
with $q_1+q_2+k=0$. Under a translation,
\begin{equation}\label{translation}
	b^\mu_{1,2}\!\to\! b^\mu_{1,2}+a^\mu,\qquad
	\tilde{\mathcal{A}}^{\mu\nu}(k)\! \to \! e^{-ia\cdot k}\!\tilde{\mathcal{A}}^{\mu\nu}(k)\,.
\end{equation}
In a frame where $b_2=0$, we find
\begin{align}\label{Awhenb2=0}
		&\tilde{\mathcal{A}}^{\mu\nu}(k) = \!\!\int\!\!\! \frac{d^Dq_1}{(2\pi)^D}\,2\pi\delta(2p_1\cdot q_1)\, e^{ib\cdot q_1}\\
		&\times 2\pi\delta(2p_2\cdot (q_1+k))\,\mathcal A^{\mu\nu}(q_1,q_2,k)\Big|_{q_2=-q_1-k}\nonumber
\end{align}
its advantage being that $k$ only enters the second line. 

\vspace{7pt}
\section{PM Impulse}
\label{app:PM}
We collect here for completeness the $\mathcal O(G)$ and $\mathcal O(G^2 c_{X_1^2})$ terms of the impulse \cite{Bern:2020uwk,Cheung:2020sdj,Kalin:2020mvi},
\begin{align}
	%	\label{Q1PM}
	\label{PM}
	Q_\text{1PM}
	&=
	\frac{4G m_1m_2}{b}\,\frac{\sigma^2-\tfrac12}{\sqrt{\sigma^2-1}}\,,
	\\
	%\label{Q2PM}
	%
	%	Q_\text{2PM}
	%	&=
	%	\frac{3\pi G^2 m_1m_2(m_1+m_2)}{4b^2}\,\frac{5\sigma^2-1}{\sqrt{\sigma^2-1}}\,,
	%	\\
	\nonumber
	%	\label{QE1}
	Q_{E_1^2}
	&=
	\frac{R_f b}{G}
	\frac{3c_{E_1^2}}{m_1^2}
	\frac{35 \sigma^4-30 \sigma^2+11}{\sqrt{\sigma ^2-1}}
	\equiv 
	\frac{R_f b}{G}\hat{Q}_{E_1^2}
	\,,
	\\
	\nonumber
	%	\label{QB1}
	Q_{B_1^2}
	&=
	\frac{R_f b}{G}
	\frac{15c_{B_1^2}}{m_1^2}
	\sqrt{\sigma ^2-1} \left(7 \sigma ^2+1\right)
	\equiv \frac{R_f b}{G} \hat{Q}_{B_1^2}\,,
\end{align}
where $R_f$ is given below Eq.~\eqref{Ptid}.

%\nocite{*}

\providecommand{\href}[2]{#2}\begingroup\raggedright\endgroup


\begin{thebibliography}{100}
	
	\bibitem{DiVecchia:2022owy}
	P.~Di~Vecchia, C.~Heissenberg, and R.~Russo, ``{Angular momentum of
		zero-frequency gravitons},''
	\href{http://dx.doi.org/10.1007/JHEP08(2022)172}{{\em JHEP} {\bf 08} (2022)
		172}, \href{http://arxiv.org/abs/2203.11915}{{\tt arXiv:2203.11915
			[hep-th]}}.
	
	\bibitem{Buonanno:2022pgc}
	A.~Buonanno, M.~Khalil, D.~O'Connell, R.~Roiban, M.~P. Solon, and M.~Zeng,
	``{Snowmass White Paper: Gravitational Waves and Scattering Amplitudes},'' in
	{\em {2022 Snowmass Summer Study}}.
	\newblock 4, 2022.
	\newblock \href{http://arxiv.org/abs/2204.05194}{{\tt arXiv:2204.05194
			[hep-th]}}.
	
	\bibitem{Note1}
	We work in natural units commonly employed in the particle physics literature.
	Therefore, we measure velocities in units of the speed of light $c$,
	effectively setting $c=1$ and identifying length and time intervals.
	Similarly, we measure actions in units of the reduced Planck constant $\hbar
	$, so that $\hbar =1$. At variance with the General Relativity literature,
	instead, we keep explicit the dependence on the Newton constant $G$, which in
	natural units has the effective dimension $(\protect \text {energy})^{-2}$.
	Of course, the appropriate powers of $c$ (and, if needed, of $\hbar $) can
	always be reinstated via dimensional analysis.
	
	\bibitem{Bjerrum-Bohr:2018xdl}
	N.~E.~J. Bjerrum-Bohr, P.~H. Damgaard, G.~Festuccia, L.~Plant{\'e}, and
	P.~Vanhove, ``{General Relativity from Scattering Amplitudes},''
	\href{http://dx.doi.org/10.1103/PhysRevLett.121.171601}{{\em Phys. Rev.
			Lett.} {\bf 121} (2018) no.~17, 171601},
	\href{http://arxiv.org/abs/1806.04920}{{\tt arXiv:1806.04920 [hep-th]}}.
	%%CITATION = ARXIV:1806.04920;%%.
	
	\bibitem{Bern:2019nnu}
	Z.~Bern, C.~Cheung, R.~Roiban, C.-H. Shen, M.~P. Solon, and M.~Zeng,
	``{Scattering Amplitudes and the Conservative Hamiltonian for Binary Systems
		at Third Post-Minkowskian Order},''
	\href{http://dx.doi.org/10.1103/PhysRevLett.122.201603}{{\em Phys. Rev.
			Lett.} {\bf 122} (2019) no.~20, 201603},
	\href{http://arxiv.org/abs/1901.04424}{{\tt arXiv:1901.04424 [hep-th]}}.
	%%CITATION = ARXIV:1901.04424;%%.
	
	\bibitem{Bern:2019crd}
	Z.~Bern, C.~Cheung, R.~Roiban, C.-H. Shen, M.~P. Solon, and M.~Zeng, ``{Black
		Hole Binary Dynamics from the Double Copy and Effective Theory},''
	\href{http://dx.doi.org/10.1007/JHEP10(2019)206}{{\em JHEP} {\bf 10} (2019)
		206}, \href{http://arxiv.org/abs/1908.01493}{{\tt arXiv:1908.01493
			[hep-th]}}.
	
	\bibitem{Bern:2021dqo}
	Z.~Bern, J.~Parra-Martinez, R.~Roiban, M.~S. Ruf, C.-H. Shen, M.~P. Solon, and
	M.~Zeng, ``{Scattering Amplitudes and Conservative Binary Dynamics at ${\cal
			O}(G^4)$},'' \href{http://dx.doi.org/10.1103/PhysRevLett.126.171601}{{\em
			Phys. Rev. Lett.} {\bf 126} (2021) no.~17, 171601},
	\href{http://arxiv.org/abs/2101.07254}{{\tt arXiv:2101.07254 [hep-th]}}.
	
	\bibitem{Bern:2021yeh}
	Z.~Bern, J.~Parra-Martinez, R.~Roiban, M.~S. Ruf, C.-H. Shen, M.~P. Solon, and
	M.~Zeng, ``{Scattering Amplitudes, the Tail Effect, and Conservative Binary
		Dynamics at O(G4)},''
	\href{http://dx.doi.org/10.1103/PhysRevLett.128.161103}{{\em Phys. Rev.
			Lett.} {\bf 128} (2022) no.~16, 161103},
	\href{http://arxiv.org/abs/2112.10750}{{\tt arXiv:2112.10750 [hep-th]}}.
	
	\bibitem{Arkani-Hamed:2017jhn}
	N.~Arkani-Hamed, T.-C. Huang, and Y.-t. Huang, ``{Scattering amplitudes for all
		masses and spins},'' \href{http://dx.doi.org/10.1007/JHEP11(2021)070}{{\em
			JHEP} {\bf 11} (2021)  070}, \href{http://arxiv.org/abs/1709.04891}{{\tt
			arXiv:1709.04891 [hep-th]}}.
	
	\bibitem{Vines:2017hyw}
	J.~Vines, ``{Scattering of two spinning black holes in post-Minkowskian
		gravity, to all orders in spin, and effective-one-body mappings},''
	\href{http://dx.doi.org/10.1088/1361-6382/aaa3a8}{{\em Class. Quant. Grav.}
		{\bf 35} (2018) no.~8, 084002}, \href{http://arxiv.org/abs/1709.06016}{{\tt
			arXiv:1709.06016 [gr-qc]}}.
	
	\bibitem{Guevara:2018wpp}
	A.~Guevara, A.~Ochirov, and J.~Vines, ``{Scattering of Spinning Black Holes
		from Exponentiated Soft Factors},''
	\href{http://dx.doi.org/10.1007/JHEP09(2019)056}{{\em JHEP} {\bf 09} (2019)
		056}, \href{http://arxiv.org/abs/1812.06895}{{\tt arXiv:1812.06895
			[hep-th]}}.
	
	\bibitem{Chung:2018kqs}
	M.-Z. Chung, Y.-T. Huang, J.-W. Kim, and S.~Lee, ``{The simplest massive
		S-matrix: from minimal coupling to Black Holes},''
	\href{http://dx.doi.org/10.1007/JHEP04(2019)156}{{\em JHEP} {\bf 04} (2019)
		156}, \href{http://arxiv.org/abs/1812.08752}{{\tt arXiv:1812.08752
			[hep-th]}}.
	
	\bibitem{Maybee:2019jus}
	B.~Maybee, D.~O'Connell, and J.~Vines, ``{Observables and amplitudes for
		spinning particles and black holes},''
	\href{http://dx.doi.org/10.1007/JHEP12(2019)156}{{\em JHEP} {\bf 12} (2019)
		156}, \href{http://arxiv.org/abs/1906.09260}{{\tt arXiv:1906.09260
			[hep-th]}}.
	
	\bibitem{Guevara:2019fsj}
	A.~Guevara, A.~Ochirov, and J.~Vines, ``{Black-hole scattering with general
		spin directions from minimal-coupling amplitudes},''
	\href{http://dx.doi.org/10.1103/PhysRevD.100.104024}{{\em Phys. Rev. D} {\bf
			100} (2019) no.~10, 104024}, \href{http://arxiv.org/abs/1906.10071}{{\tt
			arXiv:1906.10071 [hep-th]}}.
	
	\bibitem{Arkani-Hamed:2019ymq}
	N.~Arkani-Hamed, Y.-t. Huang, and D.~O'Connell, ``{Kerr black holes as
		elementary particles},''
	\href{http://dx.doi.org/10.1007/JHEP01(2020)046}{{\em JHEP} {\bf 01} (2020)
		046}, \href{http://arxiv.org/abs/1906.10100}{{\tt arXiv:1906.10100
			[hep-th]}}.
	
	\bibitem{Johansson:2019dnu}
	H.~Johansson and A.~Ochirov, ``{Double copy for massive quantum particles with
		spin},'' \href{http://dx.doi.org/10.1007/JHEP09(2019)040}{{\em JHEP} {\bf 09}
		(2019)  040}, \href{http://arxiv.org/abs/1906.12292}{{\tt arXiv:1906.12292
			[hep-th]}}.
	
	\bibitem{Chung:2019duq}
	M.-Z. Chung, Y.-T. Huang, and J.-W. Kim, ``{Classical potential for general
		spinning bodies},'' \href{http://dx.doi.org/10.1007/JHEP09(2020)074}{{\em
			JHEP} {\bf 09} (2020)  074}, \href{http://arxiv.org/abs/1908.08463}{{\tt
			arXiv:1908.08463 [hep-th]}}.
	
	\bibitem{Damgaard:2019lfh}
	P.~H. Damgaard, K.~Haddad, and A.~Helset, ``{Heavy Black Hole Effective
		Theory},'' \href{http://dx.doi.org/10.1007/JHEP11(2019)070}{{\em JHEP} {\bf
			11} (2019)  070}, \href{http://arxiv.org/abs/1908.10308}{{\tt
			arXiv:1908.10308 [hep-ph]}}.
	
	\bibitem{Bautista:2019evw}
	Y.~F. Bautista and A.~Guevara, ``{On the double copy for spinning matter},''
	\href{http://dx.doi.org/10.1007/JHEP11(2021)184}{{\em JHEP} {\bf 11} (2021)
		184}, \href{http://arxiv.org/abs/1908.11349}{{\tt arXiv:1908.11349
			[hep-th]}}.
	
	\bibitem{Aoude:2020onz}
	R.~Aoude, K.~Haddad, and A.~Helset, ``{On-shell heavy particle effective
		theories},'' \href{http://dx.doi.org/10.1007/JHEP05(2020)051}{{\em JHEP} {\bf
			05} (2020)  051}, \href{http://arxiv.org/abs/2001.09164}{{\tt
			arXiv:2001.09164 [hep-th]}}.
	
	\bibitem{Chung:2020rrz}
	M.-Z. Chung, Y.-t. Huang, J.-W. Kim, and S.~Lee, ``{Complete Hamiltonian for
		spinning binary systems at first post-Minkowskian order},''
	\href{http://dx.doi.org/10.1007/JHEP05(2020)105}{{\em JHEP} {\bf 05} (2020)
		105}, \href{http://arxiv.org/abs/2003.06600}{{\tt arXiv:2003.06600
			[hep-th]}}.
	
	\bibitem{Bern:2020buy}
	Z.~Bern, A.~Luna, R.~Roiban, C.-H. Shen, and M.~Zeng, ``{Spinning black hole
		binary dynamics, scattering amplitudes, and effective field theory},''
	\href{http://dx.doi.org/10.1103/PhysRevD.104.065014}{{\em Phys. Rev. D} {\bf
			104} (2021) no.~6, 065014}, \href{http://arxiv.org/abs/2005.03071}{{\tt
			arXiv:2005.03071 [hep-th]}}.
	
	\bibitem{Guevara:2020xjx}
	A.~Guevara, B.~Maybee, A.~Ochirov, D.~O'connell, and J.~Vines, ``{A worldsheet
		for Kerr},'' \href{http://dx.doi.org/10.1007/JHEP03(2021)201}{{\em JHEP} {\bf
			03} (2021)  201}, \href{http://arxiv.org/abs/2012.11570}{{\tt
			arXiv:2012.11570 [hep-th]}}.
	
	\bibitem{Kosmopoulos:2021zoq}
	D.~Kosmopoulos and A.~Luna, ``{Quadratic-in-spin Hamiltonian at $ \mathcal{O}
		$(G$^{2}$) from scattering amplitudes},''
	\href{http://dx.doi.org/10.1007/JHEP07(2021)037}{{\em JHEP} {\bf 07} (2021)
		037}, \href{http://arxiv.org/abs/2102.10137}{{\tt arXiv:2102.10137
			[hep-th]}}.
	
	\bibitem{Aoude:2021oqj}
	R.~Aoude and A.~Ochirov, ``{Classical observables from coherent-spin
		amplitudes},'' \href{http://dx.doi.org/10.1007/JHEP10(2021)008}{{\em JHEP}
		{\bf 10} (2021)  008}, \href{http://arxiv.org/abs/2108.01649}{{\tt
			arXiv:2108.01649 [hep-th]}}.
	
	\bibitem{Bautista:2021wfy}
	Y.~F. Bautista, A.~Guevara, C.~Kavanagh, and J.~Vines, ``{Scattering in black
		hole backgrounds and higher-spin amplitudes. Part I},''
	\href{http://dx.doi.org/10.1007/JHEP03(2023)136}{{\em JHEP} {\bf 03} (2023)
		136}, \href{http://arxiv.org/abs/2107.10179}{{\tt arXiv:2107.10179
			[hep-th]}}.
	
	\bibitem{Chiodaroli:2021eug}
	M.~Chiodaroli, H.~Johansson, and P.~Pichini, ``{Compton black-hole scattering
		for s \ensuremath{\leq} 5/2},''
	\href{http://dx.doi.org/10.1007/JHEP02(2022)156}{{\em JHEP} {\bf 02} (2022)
		156}, \href{http://arxiv.org/abs/2107.14779}{{\tt arXiv:2107.14779
			[hep-th]}}.
	
	\bibitem{Haddad:2021znf}
	K.~Haddad, ``{Exponentiation of the leading eikonal phase with spin},''
	\href{http://dx.doi.org/10.1103/PhysRevD.105.026004}{{\em Phys. Rev. D} {\bf
			105} (2022) no.~2, 026004}, \href{http://arxiv.org/abs/2109.04427}{{\tt
			arXiv:2109.04427 [hep-th]}}.
	
	\bibitem{Chen:2021kxt}
	W.-M. Chen, M.-Z. Chung, Y.-t. Huang, and J.-W. Kim, ``{The 2PM Hamiltonian for
		binary Kerr to quartic in spin},''
	\href{http://dx.doi.org/10.1007/JHEP08(2022)148}{{\em JHEP} {\bf 08} (2022)
		148}, \href{http://arxiv.org/abs/2111.13639}{{\tt arXiv:2111.13639
			[hep-th]}}.
	
	\bibitem{Aoude:2022trd}
	R.~Aoude, K.~Haddad, and A.~Helset, ``{Searching for Kerr in the 2PM
		amplitude},'' \href{http://dx.doi.org/10.1007/JHEP07(2022)072}{{\em JHEP}
		{\bf 07} (2022)  072}, \href{http://arxiv.org/abs/2203.06197}{{\tt
			arXiv:2203.06197 [hep-th]}}.
	
	\bibitem{Bern:2022kto}
	Z.~Bern, D.~Kosmopoulos, A.~Luna, R.~Roiban, and F.~Teng, ``{Binary Dynamics
		through the Fifth Power of Spin at O(G2)},''
	\href{http://dx.doi.org/10.1103/PhysRevLett.130.201402}{{\em Phys. Rev.
			Lett.} {\bf 130} (2023) no.~20, 201402},
	\href{http://arxiv.org/abs/2203.06202}{{\tt arXiv:2203.06202 [hep-th]}}.
	
	\bibitem{Alessio:2022kwv}
	F.~Alessio and P.~Di~Vecchia, ``{Radiation reaction for spinning black-hole
		scattering},'' \href{http://dx.doi.org/10.1016/j.physletb.2022.137258}{{\em
			Phys. Lett. B} {\bf 832} (2022)  137258},
	\href{http://arxiv.org/abs/2203.13272}{{\tt arXiv:2203.13272 [hep-th]}}.
	
	\bibitem{FebresCordero:2022jts}
	F.~Febres~Cordero, M.~Kraus, G.~Lin, M.~S. Ruf, and M.~Zeng, ``{Conservative
		Binary Dynamics with a Spinning Black Hole at O(G3) from Scattering
		Amplitudes},'' \href{http://dx.doi.org/10.1103/PhysRevLett.130.021601}{{\em
			Phys. Rev. Lett.} {\bf 130} (2023) no.~2, 021601},
	\href{http://arxiv.org/abs/2205.07357}{{\tt arXiv:2205.07357 [hep-th]}}.
	
	\bibitem{Cangemi:2022abk}
	L.~Cangemi and P.~Pichini, ``{Classical Limit of Higher-Spin String
		Amplitudes},'' \href{http://arxiv.org/abs/2207.03947}{{\tt arXiv:2207.03947
			[hep-th]}}.
	
	\bibitem{Cheung:2020sdj}
	C.~Cheung and M.~P. Solon, ``{Tidal Effects in the Post-Minkowskian
		Expansion},'' \href{http://dx.doi.org/10.1103/PhysRevLett.125.191601}{{\em
			Phys. Rev. Lett.} {\bf 125} (2020) no.~19, 191601},
	\href{http://arxiv.org/abs/2006.06665}{{\tt arXiv:2006.06665 [hep-th]}}.
	
	\bibitem{Bern:2020uwk}
	Z.~Bern, J.~Parra-Martinez, R.~Roiban, E.~Sawyer, and C.-H. Shen, ``{Leading
		Nonlinear Tidal Effects and Scattering Amplitudes},''
	\href{http://dx.doi.org/10.1007/JHEP05(2021)188}{{\em JHEP} {\bf 05} (2021)
		188}, \href{http://arxiv.org/abs/2010.08559}{{\tt arXiv:2010.08559
			[hep-th]}}.
	
	\bibitem{Cheung:2020gbf}
	C.~Cheung, N.~Shah, and M.~P. Solon, ``{Mining the Geodesic Equation for
		Scattering Data},'' \href{http://dx.doi.org/10.1103/PhysRevD.103.024030}{{\em
			Phys. Rev. D} {\bf 103} (2021) no.~2, 024030},
	\href{http://arxiv.org/abs/2010.08568}{{\tt arXiv:2010.08568 [hep-th]}}.
	
	\bibitem{Aoude:2020ygw}
	R.~Aoude, K.~Haddad, and A.~Helset, ``{Tidal effects for spinning particles},''
	\href{http://dx.doi.org/10.1007/JHEP03(2021)097}{{\em JHEP} {\bf 03} (2021)
		097}, \href{http://arxiv.org/abs/2012.05256}{{\tt arXiv:2012.05256
			[hep-th]}}.
	
	\bibitem{AccettulliHuber:2020dal}
	M.~Accettulli~Huber, A.~Brandhuber, S.~De~Angelis, and G.~Travaglini, ``{From
		amplitudes to gravitational radiation with cubic interactions and tidal
		effects},'' \href{http://dx.doi.org/10.1103/PhysRevD.103.045015}{{\em Phys.
			Rev. D} {\bf 103} (2021) no.~4, 045015},
	\href{http://arxiv.org/abs/2012.06548}{{\tt arXiv:2012.06548 [hep-th]}}.
	
	\bibitem{Damour:1992qi}
	T.~Damour, M.~Soffel, and C.-m. Xu, ``{General relativistic celestial
		mechanics. 3. Rotational equations of motion},''
	\href{http://dx.doi.org/10.1103/PhysRevD.47.3124}{{\em Phys. Rev. D} {\bf 47}
		(1993)  3124--3135}.
	
	\bibitem{Damour:1993zn}
	T.~Damour, M.~Soffel, and C.-m. Xu, ``{General relativistic celestial
		mechanics. 4: Theory of satellite motion},''
	\href{http://dx.doi.org/10.1103/PhysRevD.49.618}{{\em Phys. Rev. D} {\bf 49}
		(1994)  618--635}.
	
	\bibitem{Goldberger:2005cd}
	W.~D. Goldberger and I.~Z. Rothstein, ``{Dissipative effects in the worldline
		approach to black hole dynamics},''
	\href{http://dx.doi.org/10.1103/PhysRevD.73.104030}{{\em Phys. Rev. D} {\bf
			73} (2006)  104030}, \href{http://arxiv.org/abs/hep-th/0511133}{{\tt
			arXiv:hep-th/0511133}}.
	
	\bibitem{Hinderer:2007mb}
	T.~Hinderer, ``{Tidal Love numbers of neutron stars},''
	\href{http://dx.doi.org/10.1086/533487}{{\em Astrophys. J.} {\bf 677} (2008)
		1216--1220}, \href{http://arxiv.org/abs/0711.2420}{{\tt arXiv:0711.2420
			[astro-ph]}}.
	
	\bibitem{Flanagan:2007ix}
	E.~E. Flanagan and T.~Hinderer, ``{Constraining neutron star tidal Love numbers
		with gravitational wave detectors},''
	\href{http://dx.doi.org/10.1103/PhysRevD.77.021502}{{\em Phys. Rev. D} {\bf
			77} (2008)  021502}, \href{http://arxiv.org/abs/0709.1915}{{\tt
			arXiv:0709.1915 [astro-ph]}}.
	
	\bibitem{Damour:2009vw}
	T.~Damour and A.~Nagar, ``{Relativistic tidal properties of neutron stars},''
	\href{http://dx.doi.org/10.1103/PhysRevD.80.084035}{{\em Phys. Rev. D} {\bf
			80} (2009)  084035}, \href{http://arxiv.org/abs/0906.0096}{{\tt
			arXiv:0906.0096 [gr-qc]}}.
	
	\bibitem{Binnington:2009bb}
	T.~Binnington and E.~Poisson, ``{Relativistic theory of tidal Love numbers},''
	\href{http://dx.doi.org/10.1103/PhysRevD.80.084018}{{\em Phys. Rev. D} {\bf
			80} (2009)  084018}, \href{http://arxiv.org/abs/0906.1366}{{\tt
			arXiv:0906.1366 [gr-qc]}}.
	
	\bibitem{Hinderer:2009ca}
	T.~Hinderer, B.~D. Lackey, R.~N. Lang, and J.~S. Read, ``{Tidal deformability
		of neutron stars with realistic equations of state and their gravitational
		wave signatures in binary inspiral},''
	\href{http://dx.doi.org/10.1103/PhysRevD.81.123016}{{\em Phys. Rev. D} {\bf
			81} (2010)  123016}, \href{http://arxiv.org/abs/0911.3535}{{\tt
			arXiv:0911.3535 [astro-ph.HE]}}.
	
	\bibitem{Kol:2011vg}
	B.~Kol and M.~Smolkin, ``{Black hole stereotyping: Induced gravito-static
		polarization},'' \href{http://dx.doi.org/10.1007/JHEP02(2012)010}{{\em JHEP}
		{\bf 02} (2012)  010}, \href{http://arxiv.org/abs/1110.3764}{{\tt
			arXiv:1110.3764 [hep-th]}}.
	
	\bibitem{Damour:2012yf}
	T.~Damour, A.~Nagar, and L.~Villain, ``{Measurability of the tidal
		polarizability of neutron stars in late-inspiral gravitational-wave
		signals},'' \href{http://dx.doi.org/10.1103/PhysRevD.85.123007}{{\em Phys.
			Rev. D} {\bf 85} (2012)  123007}, \href{http://arxiv.org/abs/1203.4352}{{\tt
			arXiv:1203.4352 [gr-qc]}}.
	
	\bibitem{Favata:2013rwa}
	M.~Favata, ``{Systematic parameter errors in inspiraling neutron star
		binaries},'' \href{http://dx.doi.org/10.1103/PhysRevLett.112.101101}{{\em
			Phys. Rev. Lett.} {\bf 112} (2014)  101101},
	\href{http://arxiv.org/abs/1310.8288}{{\tt arXiv:1310.8288 [gr-qc]}}.
	
	\bibitem{Baiotti:2016qnr}
	L.~Baiotti and L.~Rezzolla, ``{Binary neutron star mergers: a review of
		Einstein\textquoteright{}s richest laboratory},''
	\href{http://dx.doi.org/10.1088/1361-6633/aa67bb}{{\em Rept. Prog. Phys.}
		{\bf 80} (2017) no.~9, 096901}, \href{http://arxiv.org/abs/1607.03540}{{\tt
			arXiv:1607.03540 [gr-qc]}}.
	
	\bibitem{Barack:2018yly}
	L.~Barack {\em et al.}, ``{Black holes, gravitational waves and fundamental
		physics: a roadmap},'' \href{http://dx.doi.org/10.1088/1361-6382/ab0587}{{\em
			Class. Quant. Grav.} {\bf 36} (2019) no.~14, 143001},
	\href{http://arxiv.org/abs/1806.05195}{{\tt arXiv:1806.05195 [gr-qc]}}.
	
	\bibitem{Buonanno:2014aza}
	A.~Buonanno and B.~S. Sathyaprakash, {\em {Sources of Gravitational Waves:
			Theory and Observations}}.
	\newblock 10, 2014.
	\newblock \href{http://arxiv.org/abs/1410.7832}{{\tt arXiv:1410.7832 [gr-qc]}}.
	
	\bibitem{Cardoso:2019rvt}
	V.~Cardoso and P.~Pani, ``{Testing the nature of dark compact objects: a status
		report},'' \href{http://dx.doi.org/10.1007/s41114-019-0020-4}{{\em Living
			Rev. Rel.} {\bf 22} (2019) no.~1, 4},
	\href{http://arxiv.org/abs/1904.05363}{{\tt arXiv:1904.05363 [gr-qc]}}.
	
	\bibitem{Baumann:2019ztm}
	D.~Baumann, H.~S. Chia, R.~A. Porto, and J.~Stout, ``{Gravitational Collider
		Physics},'' \href{http://dx.doi.org/10.1103/PhysRevD.101.083019}{{\em Phys.
			Rev. D} {\bf 101} (2020) no.~8, 083019},
	\href{http://arxiv.org/abs/1912.04932}{{\tt arXiv:1912.04932 [gr-qc]}}.
	
	\bibitem{Bern:2008qj}
	Z.~Bern, J.~J.~M. Carrasco, and H.~Johansson, ``{New Relations for Gauge-Theory
		Amplitudes},'' \href{http://dx.doi.org/10.1103/PhysRevD.78.085011}{{\em Phys.
			Rev. D} {\bf 78} (2008)  085011}, \href{http://arxiv.org/abs/0805.3993}{{\tt
			arXiv:0805.3993 [hep-ph]}}.
	
	\bibitem{Bern:2010ue}
	Z.~Bern, J.~J.~M. Carrasco, and H.~Johansson, ``{Perturbative Quantum Gravity
		as a Double Copy of Gauge Theory},''
	\href{http://dx.doi.org/10.1103/PhysRevLett.105.061602}{{\em Phys. Rev.
			Lett.} {\bf 105} (2010)  061602}, \href{http://arxiv.org/abs/1004.0476}{{\tt
			arXiv:1004.0476 [hep-th]}}.
	
	\bibitem{Bern:2012uf}
	Z.~Bern, J.~J.~M. Carrasco, L.~J. Dixon, H.~Johansson, and R.~Roiban,
	``{Simplifying Multiloop Integrands and Ultraviolet Divergences of Gauge
		Theory and Gravity Amplitudes},''
	\href{http://dx.doi.org/10.1103/PhysRevD.85.105014}{{\em Phys. Rev. D} {\bf
			85} (2012)  105014}, \href{http://arxiv.org/abs/1201.5366}{{\tt
			arXiv:1201.5366 [hep-th]}}.
	
	\bibitem{Bern:2017ucb}
	Z.~Bern, J.~J.~M. Carrasco, W.-M. Chen, H.~Johansson, R.~Roiban, and M.~Zeng,
	``{Five-loop four-point integrand of $N=8$ supergravity as a generalized
		double copy},'' \href{http://dx.doi.org/10.1103/PhysRevD.96.126012}{{\em
			Phys. Rev. D} {\bf 96} (2017) no.~12, 126012},
	\href{http://arxiv.org/abs/1708.06807}{{\tt arXiv:1708.06807 [hep-th]}}.
	
	\bibitem{Bern:2018jmv}
	Z.~Bern, J.~J. Carrasco, W.-M. Chen, A.~Edison, H.~Johansson,
	J.~Parra-Martinez, R.~Roiban, and M.~Zeng, ``{Ultraviolet Properties of
		$\mathcal N = 8$ Supergravity at Five Loops},''
	\href{http://dx.doi.org/10.1103/PhysRevD.98.086021}{{\em Phys. Rev. D} {\bf
			98} (2018) no.~8, 086021}, \href{http://arxiv.org/abs/1804.09311}{{\tt
			arXiv:1804.09311 [hep-th]}}.
	
	\bibitem{Bern:2019prr}
	Z.~Bern, J.~J. Carrasco, M.~Chiodaroli, H.~Johansson, and R.~Roiban, ``{The
		Duality Between Color and Kinematics and its Applications},''
	\href{http://arxiv.org/abs/1909.01358}{{\tt arXiv:1909.01358 [hep-th]}}.
	
	\bibitem{Bern:1994zx}
	Z.~Bern, L.~J. Dixon, D.~C. Dunbar, and D.~A. Kosower, ``{One loop n point
		gauge theory amplitudes, unitarity and collinear limits},''
	\href{http://dx.doi.org/10.1016/0550-3213(94)90179-1}{{\em Nucl. Phys. B}
		{\bf 425} (1994)  217--260}, \href{http://arxiv.org/abs/hep-ph/9403226}{{\tt
			arXiv:hep-ph/9403226}}.
	
	\bibitem{Bern:1994cg}
	Z.~Bern, L.~J. Dixon, D.~C. Dunbar, and D.~A. Kosower, ``{Fusing gauge theory
		tree amplitudes into loop amplitudes},''
	\href{http://dx.doi.org/10.1016/0550-3213(94)00488-Z}{{\em Nucl. Phys. B}
		{\bf 435} (1995)  59--101}, \href{http://arxiv.org/abs/hep-ph/9409265}{{\tt
			arXiv:hep-ph/9409265}}.
	
	\bibitem{Britto:2004nc}
	R.~Britto, F.~Cachazo, and B.~Feng, ``{Generalized unitarity and one-loop
		amplitudes in N=4 super-Yang-Mills},''
	\href{http://dx.doi.org/10.1016/j.nuclphysb.2005.07.014}{{\em Nucl. Phys. B}
		{\bf 725} (2005)  275--305}, \href{http://arxiv.org/abs/hep-th/0412103}{{\tt
			arXiv:hep-th/0412103}}.
	
	\bibitem{Kabat:1992tb}
	D.~N. Kabat and M.~Ortiz, ``{Eikonal quantum gravity and Planckian
		scattering},'' \href{http://dx.doi.org/10.1016/0550-3213(92)90627-N}{{\em
			Nucl. Phys.} {\bf B388} (1992)  570--592},
	\href{http://arxiv.org/abs/hep-th/9203082}{{\tt arXiv:hep-th/9203082
			[hep-th]}}.
	%%CITATION = HEP-TH/9203082;%%.
	
	\bibitem{Akhoury:2013yua}
	R.~Akhoury, R.~Saotome, and G.~Sterman, ``{High Energy Scattering in
		Perturbative Quantum Gravity at Next to Leading Power},''
	\href{http://dx.doi.org/10.1103/PhysRevD.103.064036}{{\em Phys. Rev. D} {\bf
			103} (2021) no.~6, 064036}, \href{http://arxiv.org/abs/1308.5204}{{\tt
			arXiv:1308.5204 [hep-th]}}.
	
	\bibitem{KoemansCollado:2019ggb}
	A.~Koemans~Collado, P.~Di~Vecchia, and R.~Russo, ``{Revisiting the second
		post-Minkowskian eikonal and the dynamics of binary black holes},''
	\href{http://dx.doi.org/10.1103/PhysRevD.100.066028}{{\em Phys. Rev. D} {\bf
			100} (2019) no.~6, 066028}, \href{http://arxiv.org/abs/1904.02667}{{\tt
			arXiv:1904.02667 [hep-th]}}.
	
	\bibitem{Cheung:2020gyp}
	C.~Cheung and M.~P. Solon, ``{Classical gravitational scattering at $
		\mathcal{O} $(G$^{3}$) from Feynman diagrams},''
	\href{http://dx.doi.org/10.1007/JHEP06(2020)144}{{\em JHEP} {\bf 06} (2020)
		144}, \href{http://arxiv.org/abs/2003.08351}{{\tt arXiv:2003.08351
			[hep-th]}}.
	
	\bibitem{DiVecchia:2020ymx}
	P.~Di~Vecchia, C.~Heissenberg, R.~Russo, and G.~Veneziano, ``{Universality of
		ultra-relativistic gravitational scattering},''
	\href{http://dx.doi.org/10.1016/j.physletb.2020.135924}{{\em Phys. Lett. B}
		{\bf 811} (2020)  135924}, \href{http://arxiv.org/abs/2008.12743}{{\tt
			arXiv:2008.12743 [hep-th]}}.
	
	\bibitem{AccettulliHuber:2020oou}
	M.~Accettulli~Huber, A.~Brandhuber, S.~De~Angelis, and G.~Travaglini,
	``{Eikonal phase matrix, deflection angle and time delay in effective field
		theories of gravity},''
	\href{http://dx.doi.org/10.1103/PhysRevD.102.046014}{{\em Phys. Rev. D} {\bf
			102} (2020) no.~4, 046014}, \href{http://arxiv.org/abs/2006.02375}{{\tt
			arXiv:2006.02375 [hep-th]}}.
	
	\bibitem{DiVecchia:2021ndb}
	P.~Di~Vecchia, C.~Heissenberg, R.~Russo, and G.~Veneziano, ``{Radiation
		Reaction from Soft Theorems},''
	\href{http://dx.doi.org/10.1016/j.physletb.2021.136379}{{\em Phys. Lett. B}
		{\bf 818} (2021)  136379}, \href{http://arxiv.org/abs/2101.05772}{{\tt
			arXiv:2101.05772 [hep-th]}}.
	
	\bibitem{DiVecchia:2021bdo}
	P.~Di~Vecchia, C.~Heissenberg, R.~Russo, and G.~Veneziano, ``{The eikonal
		approach to gravitational scattering and radiation at $ \mathcal{O}
		$(G$^{3}$)},'' \href{http://dx.doi.org/10.1007/JHEP07(2021)169}{{\em JHEP}
		{\bf 07} (2021)  169}, \href{http://arxiv.org/abs/2104.03256}{{\tt
			arXiv:2104.03256 [hep-th]}}.
	
	\bibitem{Heissenberg:2021tzo}
	C.~Heissenberg, ``{Infrared divergences and the eikonal exponentiation},''
	\href{http://dx.doi.org/10.1103/PhysRevD.104.046016}{{\em Phys. Rev. D} {\bf
			104} (2021) no.~4, 046016}, \href{http://arxiv.org/abs/2105.04594}{{\tt
			arXiv:2105.04594 [hep-th]}}.
	
	\bibitem{Bjerrum-Bohr:2021vuf}
	N.~E.~J. Bjerrum-Bohr, P.~H. Damgaard, L.~Plant\'e, and P.~Vanhove,
	``{Classical gravity from loop amplitudes},''
	\href{http://dx.doi.org/10.1103/PhysRevD.104.026009}{{\em Phys. Rev. D} {\bf
			104} (2021) no.~2, 026009}, \href{http://arxiv.org/abs/2104.04510}{{\tt
			arXiv:2104.04510 [hep-th]}}.
	
	\bibitem{Bjerrum-Bohr:2021din}
	N.~E.~J. Bjerrum-Bohr, P.~H. Damgaard, L.~Plant\'e, and P.~Vanhove, ``{The
		amplitude for classical gravitational scattering at third Post-Minkowskian
		order},'' \href{http://dx.doi.org/10.1007/JHEP08(2021)172}{{\em JHEP} {\bf
			08} (2021)  172}, \href{http://arxiv.org/abs/2105.05218}{{\tt
			arXiv:2105.05218 [hep-th]}}.
	
	\bibitem{Damgaard:2021ipf}
	P.~H. Damgaard, L.~Plante, and P.~Vanhove, ``{On an exponential representation
		of the gravitational S-matrix},''
	\href{http://dx.doi.org/10.1007/JHEP11(2021)213}{{\em JHEP} {\bf 11} (2021)
		213}, \href{http://arxiv.org/abs/2107.12891}{{\tt arXiv:2107.12891
			[hep-th]}}.
	
	\bibitem{Brandhuber:2021eyq}
	A.~Brandhuber, G.~Chen, G.~Travaglini, and C.~Wen, ``{Classical gravitational
		scattering from a gauge-invariant double copy},''
	\href{http://dx.doi.org/10.1007/JHEP10(2021)118}{{\em JHEP} {\bf 10} (2021)
		118}, \href{http://arxiv.org/abs/2108.04216}{{\tt arXiv:2108.04216
			[hep-th]}}.
	
	\bibitem{DiVecchia:2022nna}
	P.~Di~Vecchia, C.~Heissenberg, R.~Russo, and G.~Veneziano, ``{The eikonal
		operator at arbitrary velocities I: the soft-radiation limit},''
	\href{http://dx.doi.org/10.1007/JHEP07(2022)039}{{\em JHEP} {\bf 07} (2022)
		039}, \href{http://arxiv.org/abs/2204.02378}{{\tt arXiv:2204.02378
			[hep-th]}}.
	
	\bibitem{Goldberger:2004jt}
	W.~D. Goldberger and I.~Z. Rothstein, ``{An Effective field theory of gravity
		for extended objects},''
	\href{http://dx.doi.org/10.1103/PhysRevD.73.104029}{{\em Phys. Rev. D} {\bf
			73} (2006)  104029}, \href{http://arxiv.org/abs/hep-th/0409156}{{\tt
			arXiv:hep-th/0409156}}.
	
	\bibitem{Porto:2016pyg}
	R.~A. Porto, ``{The effective field theorist\textquoteright{}s approach to
		gravitational dynamics},''
	\href{http://dx.doi.org/10.1016/j.physrep.2016.04.003}{{\em Phys. Rept.} {\bf
			633} (2016)  1--104}, \href{http://arxiv.org/abs/1601.04914}{{\tt
			arXiv:1601.04914 [hep-th]}}.
	
	\bibitem{Cheung:2018wkq}
	C.~Cheung, I.~Z. Rothstein, and M.~P. Solon, ``{From Scattering Amplitudes to
		Classical Potentials in the Post-Minkowskian Expansion},''
	\href{http://dx.doi.org/10.1103/PhysRevLett.121.251101}{{\em Phys. Rev.
			Lett.} {\bf 121} (2018) no.~25, 251101},
	\href{http://arxiv.org/abs/1808.02489}{{\tt arXiv:1808.02489 [hep-th]}}.
	%%CITATION = ARXIV:1808.02489;%%.
	
	\bibitem{Cristofoli:2020uzm}
	A.~Cristofoli, P.~H. Damgaard, P.~Di~Vecchia, and C.~Heissenberg,
	``{Second-order Post-Minkowskian scattering in arbitrary dimensions},''
	\href{http://dx.doi.org/10.1007/JHEP07(2020)122}{{\em JHEP} {\bf 07} (2020)
		122}, \href{http://arxiv.org/abs/2003.10274}{{\tt arXiv:2003.10274
			[hep-th]}}.
	
	\bibitem{Kosower:2018adc}
	D.~A. Kosower, B.~Maybee, and D.~O'Connell, ``{Amplitudes, Observables, and
		Classical Scattering},''
	\href{http://dx.doi.org/10.1007/JHEP02(2019)137}{{\em JHEP} {\bf 02} (2019)
		137},
	\href{http://arxiv.org/abs/1811.10950}{{\tt arXiv:1811.10950 [hep-th]}}.
	%%CITATION = ARXIV:1811.10950;%%.
	
	\bibitem{Herrmann:2021lqe}
	E.~Herrmann, J.~Parra-Martinez, M.~S. Ruf, and M.~Zeng, ``{Gravitational
		Bremsstrahlung from Reverse Unitarity},''
	\href{http://dx.doi.org/10.1103/PhysRevLett.126.201602}{{\em Phys. Rev.
			Lett.} {\bf 126} (2021) no.~20, 201602},
	\href{http://arxiv.org/abs/2101.07255}{{\tt arXiv:2101.07255 [hep-th]}}.
	
	\bibitem{Herrmann:2021tct}
	E.~Herrmann, J.~Parra-Martinez, M.~S. Ruf, and M.~Zeng, ``{Radiative classical
		gravitational observables at $ \mathcal{O} $(G$^{3}$) from scattering
		amplitudes},'' \href{http://dx.doi.org/10.1007/JHEP10(2021)148}{{\em JHEP}
		{\bf 10} (2021)  148}, \href{http://arxiv.org/abs/2104.03957}{{\tt
			arXiv:2104.03957 [hep-th]}}.
	
	\bibitem{Cristofoli:2021vyo}
	A.~Cristofoli, R.~Gonzo, D.~A. Kosower, and D.~O'Connell, ``{Waveforms from
		amplitudes},'' \href{http://dx.doi.org/10.1103/PhysRevD.106.056007}{{\em
			Phys. Rev. D} {\bf 106} (2022) no.~5, 056007},
	\href{http://arxiv.org/abs/2107.10193}{{\tt arXiv:2107.10193 [hep-th]}}.
	
	\bibitem{Cristofoli:2021jas}
	A.~Cristofoli, R.~Gonzo, N.~Moynihan, D.~O'Connell, A.~Ross, M.~Sergola, and
	C.~D. White, ``{The Uncertainty Principle and Classical Amplitudes},''
	\href{http://arxiv.org/abs/2112.07556}{{\tt arXiv:2112.07556 [hep-th]}}.
	
	\bibitem{Adamo:2022rmp}
	T.~Adamo, A.~Cristofoli, and A.~Ilderton, ``{Classical physics from amplitudes
		on curved backgrounds},''
	\href{http://dx.doi.org/10.1007/JHEP08(2022)281}{{\em JHEP} {\bf 08} (2022)
		281}, \href{http://arxiv.org/abs/2203.13785}{{\tt arXiv:2203.13785
			[hep-th]}}.
	
	\bibitem{Adamo:2022qci}
	T.~Adamo, A.~Cristofoli, A.~Ilderton, and S.~Klisch, ``{All Order Gravitational
		Waveforms from Scattering Amplitudes},''
	\href{http://dx.doi.org/10.1103/PhysRevLett.131.011601}{{\em Phys. Rev.
			Lett.} {\bf 131} (2023) no.~1, 011601},
	\href{http://arxiv.org/abs/2210.04696}{{\tt arXiv:2210.04696 [hep-th]}}.
	
	\bibitem{Parra-Martinez:2020dzs}
	J.~Parra-Martinez, M.~S. Ruf, and M.~Zeng, ``{Extremal black hole scattering at
		$\mathcal{O}(G^3)$: graviton dominance, eikonal exponentiation, and
		differential equations},''
	\href{http://dx.doi.org/10.1007/JHEP11(2020)023}{{\em JHEP} {\bf 11} (2020)
		023}, \href{http://arxiv.org/abs/2005.04236}{{\tt arXiv:2005.04236
			[hep-th]}}.
	
	\bibitem{Anastasiou:2002qz}
	C.~Anastasiou, L.~J. Dixon, and K.~Melnikov, ``{NLO Higgs boson rapidity
		distributions at hadron colliders},''
	\href{http://dx.doi.org/10.1016/S0920-5632(03)80168-8}{{\em Nucl. Phys. B
			Proc. Suppl.} {\bf 116} (2003)  193--197},
	\href{http://arxiv.org/abs/hep-ph/0211141}{{\tt arXiv:hep-ph/0211141}}.
	
	\bibitem{Anastasiou:2002yz}
	C.~Anastasiou and K.~Melnikov, ``{Higgs boson production at hadron colliders in
		NNLO QCD},'' \href{http://dx.doi.org/10.1016/S0550-3213(02)00837-4}{{\em
			Nucl. Phys. B} {\bf 646} (2002)  220--256},
	\href{http://arxiv.org/abs/hep-ph/0207004}{{\tt arXiv:hep-ph/0207004}}.
	
	\bibitem{Anastasiou:2003yy}
	C.~Anastasiou, L.~J. Dixon, K.~Melnikov, and F.~Petriello, ``{Dilepton rapidity
		distribution in the Drell-Yan process at NNLO in QCD},''
	\href{http://dx.doi.org/10.1103/PhysRevLett.91.182002}{{\em Phys. Rev. Lett.}
		{\bf 91} (2003)  182002}, \href{http://arxiv.org/abs/hep-ph/0306192}{{\tt
			arXiv:hep-ph/0306192}}.
	
	\bibitem{Anastasiou:2015yha}
	C.~Anastasiou, C.~Duhr, F.~Dulat, E.~Furlan, F.~Herzog, and B.~Mistlberger,
	``{Soft expansion of double-real-virtual corrections to Higgs production at
		N$^{3}$LO},'' \href{http://dx.doi.org/10.1007/JHEP08(2015)051}{{\em JHEP}
		{\bf 08} (2015)  051}, \href{http://arxiv.org/abs/1505.04110}{{\tt
			arXiv:1505.04110 [hep-ph]}}.
	
	\bibitem{Goldberger:2016iau}
	W.~D. Goldberger and A.~K. Ridgway, ``{Radiation and the classical double copy
		for color charges},''
	\href{http://dx.doi.org/10.1103/PhysRevD.95.125010}{{\em Phys. Rev. D} {\bf
			95} (2017) no.~12, 125010}, \href{http://arxiv.org/abs/1611.03493}{{\tt
			arXiv:1611.03493 [hep-th]}}.
	
	\bibitem{Kalin:2020mvi}
	G.~K\"alin and R.~A. Porto, ``{Post-Minkowskian Effective Field Theory for
		Conservative Binary Dynamics},''
	\href{http://dx.doi.org/10.1007/JHEP11(2020)106}{{\em JHEP} {\bf 11} (2020)
		106}, \href{http://arxiv.org/abs/2006.01184}{{\tt arXiv:2006.01184
			[hep-th]}}.
	
	\bibitem{Kalin:2020fhe}
	G.~K\"alin, Z.~Liu, and R.~A. Porto, ``{Conservative Dynamics of Binary Systems
		to Third Post-Minkowskian Order from the Effective Field Theory Approach},''
	\href{http://dx.doi.org/10.1103/PhysRevLett.125.261103}{{\em Phys. Rev.
			Lett.} {\bf 125} (2020) no.~26, 261103},
	\href{http://arxiv.org/abs/2007.04977}{{\tt arXiv:2007.04977 [hep-th]}}.
	
	\bibitem{Kalin:2020lmz}
	G.~K\"alin, Z.~Liu, and R.~A. Porto, ``{Conservative Tidal Effects in Compact
		Binary Systems to Next-to-Leading Post-Minkowskian Order},''
	\href{http://dx.doi.org/10.1103/PhysRevD.102.124025}{{\em Phys. Rev. D} {\bf
			102} (2020)  124025}, \href{http://arxiv.org/abs/2008.06047}{{\tt
			arXiv:2008.06047 [hep-th]}}.
	
	\bibitem{Mogull:2020sak}
	G.~Mogull, J.~Plefka, and J.~Steinhoff, ``{Classical black hole scattering from
		a worldline quantum field theory},''
	\href{http://dx.doi.org/10.1007/JHEP02(2021)048}{{\em JHEP} {\bf 02} (2021)
		048}, \href{http://arxiv.org/abs/2010.02865}{{\tt arXiv:2010.02865
			[hep-th]}}.
	
	\bibitem{Jakobsen:2021smu}
	G.~U. Jakobsen, G.~Mogull, J.~Plefka, and J.~Steinhoff, ``{Classical
		Gravitational Bremsstrahlung from a Worldline Quantum Field Theory},''
	\href{http://dx.doi.org/10.1103/PhysRevLett.126.201103}{{\em Phys. Rev.
			Lett.} {\bf 126} (2021) no.~20, 201103},
	\href{http://arxiv.org/abs/2101.12688}{{\tt arXiv:2101.12688 [gr-qc]}}.
	
	\bibitem{Mougiakakos:2021ckm}
	S.~Mougiakakos, M.~M. Riva, and F.~Vernizzi, ``{Gravitational Bremsstrahlung in
		the post-Minkowskian effective field theory},''
	\href{http://dx.doi.org/10.1103/PhysRevD.104.024041}{{\em Phys. Rev. D} {\bf
			104} (2021) no.~2, 024041}, \href{http://arxiv.org/abs/2102.08339}{{\tt
			arXiv:2102.08339 [gr-qc]}}.
	
	\bibitem{Liu:2021zxr}
	Z.~Liu, R.~A. Porto, and Z.~Yang, ``{Spin Effects in the Effective Field Theory
		Approach to Post-Minkowskian Conservative Dynamics},''
	\href{http://dx.doi.org/10.1007/JHEP06(2021)012}{{\em JHEP} {\bf 06} (2021)
		012}, \href{http://arxiv.org/abs/2102.10059}{{\tt arXiv:2102.10059
			[hep-th]}}.
	
	\bibitem{Dlapa:2021npj}
	C.~Dlapa, G.~K\"alin, Z.~Liu, and R.~A. Porto, ``{Dynamics of binary systems to
		fourth Post-Minkowskian order from the effective field theory approach},''
	\href{http://dx.doi.org/10.1016/j.physletb.2022.137203}{{\em Phys. Lett. B}
		{\bf 831} (2022)  137203}, \href{http://arxiv.org/abs/2106.08276}{{\tt
			arXiv:2106.08276 [hep-th]}}.
	
	\bibitem{Jakobsen:2021lvp}
	G.~U. Jakobsen, G.~Mogull, J.~Plefka, and J.~Steinhoff, ``{Gravitational
		Bremsstrahlung and Hidden Supersymmetry of Spinning Bodies},''
	\href{http://dx.doi.org/10.1103/PhysRevLett.128.011101}{{\em Phys. Rev.
			Lett.} {\bf 128} (2022) no.~1, 011101},
	\href{http://arxiv.org/abs/2106.10256}{{\tt arXiv:2106.10256 [hep-th]}}.
	
	\bibitem{Jakobsen:2021zvh}
	G.~U. Jakobsen, G.~Mogull, J.~Plefka, and J.~Steinhoff, ``{SUSY in the sky with
		gravitons},'' \href{http://dx.doi.org/10.1007/JHEP01(2022)027}{{\em JHEP}
		{\bf 01} (2022)  027}, \href{http://arxiv.org/abs/2109.04465}{{\tt
			arXiv:2109.04465 [hep-th]}}.
	
	\bibitem{Riva:2021vnj}
	M.~M. Riva and F.~Vernizzi, ``{Radiated momentum in the post-Minkowskian
		worldline approach via reverse unitarity},''
	\href{http://dx.doi.org/10.1007/JHEP11(2021)228}{{\em JHEP} {\bf 11} (2021)
		228}, \href{http://arxiv.org/abs/2110.10140}{{\tt arXiv:2110.10140
			[hep-th]}}.
	
	\bibitem{Dlapa:2021vgp}
	C.~Dlapa, G.~K\"alin, Z.~Liu, and R.~A. Porto, ``{Conservative Dynamics of
		Binary Systems at Fourth Post-Minkowskian Order in the Large-Eccentricity
		Expansion},'' \href{http://dx.doi.org/10.1103/PhysRevLett.128.161104}{{\em
			Phys. Rev. Lett.} {\bf 128} (2022) no.~16, 161104},
	\href{http://arxiv.org/abs/2112.11296}{{\tt arXiv:2112.11296 [hep-th]}}.
	
	\bibitem{Jakobsen:2022fcj}
	G.~U. Jakobsen and G.~Mogull, ``{Conservative and Radiative Dynamics of
		Spinning Bodies at Third Post-Minkowskian Order Using Worldline Quantum Field
		Theory},'' \href{http://dx.doi.org/10.1103/PhysRevLett.128.141102}{{\em Phys.
			Rev. Lett.} {\bf 128} (2022) no.~14, 141102},
	\href{http://arxiv.org/abs/2201.07778}{{\tt arXiv:2201.07778 [hep-th]}}.
	
	\bibitem{Mougiakakos:2022sic}
	S.~Mougiakakos, M.~M. Riva, and F.~Vernizzi, ``{Gravitational Bremsstrahlung
		with Tidal Effects in the Post-Minkowskian Expansion},''
	\href{http://dx.doi.org/10.1103/PhysRevLett.129.121101}{{\em Phys. Rev.
			Lett.} {\bf 129} (2022) no.~12, 121101},
	\href{http://arxiv.org/abs/2204.06556}{{\tt arXiv:2204.06556 [hep-th]}}.
	
	\bibitem{Jakobsen:2022psy}
	G.~U. Jakobsen, G.~Mogull, J.~Plefka, and B.~Sauer, ``{All things retarded:
		radiation-reaction in worldline quantum field theory},''
	\href{http://dx.doi.org/10.1007/JHEP10(2022)128}{{\em JHEP} {\bf 10} (2022)
		128}, \href{http://arxiv.org/abs/2207.00569}{{\tt arXiv:2207.00569
			[hep-th]}}.
	
	\bibitem{Kalin:2022hph}
	G.~K\"alin, J.~Neef, and R.~A. Porto, ``{Radiation-reaction in the Effective
		Field Theory approach to Post-Minkowskian dynamics},''
	\href{http://dx.doi.org/10.1007/JHEP01(2023)140}{{\em JHEP} {\bf 01} (2023)
		140}, \href{http://arxiv.org/abs/2207.00580}{{\tt arXiv:2207.00580
			[hep-th]}}.
	
	\bibitem{Dlapa:2022lmu}
	C.~Dlapa, G.~K\"alin, Z.~Liu, J.~Neef, and R.~A. Porto, ``{Radiation Reaction
		and Gravitational Waves at Fourth Post-Minkowskian Order},''
	\href{http://dx.doi.org/10.1103/PhysRevLett.130.101401}{{\em Phys. Rev.
			Lett.} {\bf 130} (2023) no.~10, 101401},
	\href{http://arxiv.org/abs/2210.05541}{{\tt arXiv:2210.05541 [hep-th]}}.
	
	\bibitem{Jakobsen:2022zsx}
	G.~U. Jakobsen and G.~Mogull, ``{Linear response, Hamiltonian, and radiative
		spinning two-body dynamics},''
	\href{http://dx.doi.org/10.1103/PhysRevD.107.044033}{{\em Phys. Rev. D} {\bf
			107} (2023) no.~4, 044033}, \href{http://arxiv.org/abs/2210.06451}{{\tt
			arXiv:2210.06451 [hep-th]}}.
	
	\bibitem{Ciafaloni:2018uwe}
	M.~Ciafaloni, D.~Colferai, and G.~Veneziano, ``{Infrared features of
		gravitational scattering and radiation in the eikonal approach},''
	\href{http://dx.doi.org/10.1103/PhysRevD.99.066008}{{\em Phys. Rev.} {\bf
			D99} (2019) no.~6, 066008},
	\href{http://arxiv.org/abs/1812.08137}{{\tt arXiv:1812.08137 [hep-th]}}.
	%%CITATION = ARXIV:1812.08137;%%.
	
	\bibitem{Addazi:2019mjh}
	A.~Addazi, M.~Bianchi, and G.~Veneziano, ``{Soft gravitational radiation from
		ultra-relativistic collisions at sub- and sub-sub-leading order},''
	\href{http://dx.doi.org/10.1007/JHEP05(2019)050}{{\em JHEP} {\bf 05} (2019)
		050},
	\href{http://arxiv.org/abs/1901.10986}{{\tt arXiv:1901.10986 [hep-th]}}.
	%%CITATION = ARXIV:1901.10986;%%.
	
	\bibitem{DiVecchia:2022piu}
	P.~Di~Vecchia, C.~Heissenberg, R.~Russo, and G.~Veneziano, ``{Classical
		gravitational observables from the Eikonal operator},''
	\href{http://dx.doi.org/10.1016/j.physletb.2023.138049}{{\em Phys. Lett. B}
		{\bf 843} (2023)  138049}, \href{http://arxiv.org/abs/2210.12118}{{\tt
			arXiv:2210.12118 [hep-th]}}.
	
	\bibitem{Manohar:2022dea}
	A.~V. Manohar, A.~K. Ridgway, and C.-H. Shen, ``{Radiated Angular Momentum and
		Dissipative Effects in Classical Scattering},''
	\href{http://dx.doi.org/10.1103/PhysRevLett.129.121601}{{\em Phys. Rev.
			Lett.} {\bf 129} (2022) no.~12, 121601},
	\href{http://arxiv.org/abs/2203.04283}{{\tt arXiv:2203.04283 [hep-th]}}.
	
	\bibitem{Damour:2008yg}
	T.~Damour, ``{Introductory lectures on the Effective One Body formalism},''
	\href{http://dx.doi.org/10.1142/S0217751X08039992}{{\em Int. J. Mod. Phys. A}
		{\bf 23} (2008)  1130--1148}, \href{http://arxiv.org/abs/0802.4047}{{\tt
			arXiv:0802.4047 [gr-qc]}}.
	
	\bibitem{Antonelli:2019ytb}
	A.~Antonelli, A.~Buonanno, J.~Steinhoff, M.~van~de Meent, and J.~Vines,
	``{Energetics of two-body Hamiltonians in post-Minkowskian gravity},''
	\href{http://dx.doi.org/10.1103/PhysRevD.99.104004}{{\em Phys. Rev. D} {\bf
			99} (2019) no.~10, 104004}, \href{http://arxiv.org/abs/1901.07102}{{\tt
			arXiv:1901.07102 [gr-qc]}}.
	
	\bibitem{Khalil:2022ylj}
	M.~Khalil, A.~Buonanno, J.~Steinhoff, and J.~Vines, ``{Energetics and
		scattering of gravitational two-body systems at fourth post-Minkowskian
		order},'' \href{http://dx.doi.org/10.1103/PhysRevD.106.024042}{{\em Phys.
			Rev. D} {\bf 106} (2022) no.~2, 024042},
	\href{http://arxiv.org/abs/2204.05047}{{\tt arXiv:2204.05047 [gr-qc]}}.
	
	\bibitem{Jakobsen:2020ksu}
	G.~U. Jakobsen, ``{Schwarzschild-Tangherlini Metric from Scattering
		Amplitudes},'' \href{http://dx.doi.org/10.1103/PhysRevD.102.104065}{{\em
			Phys. Rev. D} {\bf 102} (2020) no.~10, 104065},
	\href{http://arxiv.org/abs/2006.01734}{{\tt arXiv:2006.01734 [hep-th]}}.
	
	\bibitem{Mougiakakos:2020laz}
	S.~Mougiakakos and P.~Vanhove, ``{Schwarzschild-Tangherlini metric from
		scattering amplitudes in various dimensions},''
	\href{http://dx.doi.org/10.1103/PhysRevD.103.026001}{{\em Phys. Rev. D} {\bf
			103} (2021) no.~2, 026001}, \href{http://arxiv.org/abs/2010.08882}{{\tt
			arXiv:2010.08882 [hep-th]}}.
	
	\bibitem{Note2}
	The more precise form of the eikonal operator \cite
	{Cristofoli:2021jas,DiVecchia:2022piu} also involves integrations over the
	massive particles' phase space, but this will not be relevant for the present
	calculations.
	
	\bibitem{Mirbabayi:2016axw}
	M.~Mirbabayi and M.~Porrati, ``{Dressed Hard States and Black Hole Soft
		Hair},'' \href{http://dx.doi.org/10.1103/PhysRevLett.117.211301}{{\em Phys.
			Rev. Lett.} {\bf 117} (2016) no.~21, 211301},
	\href{http://arxiv.org/abs/1607.03120}{{\tt arXiv:1607.03120 [hep-th]}}.
	
	\bibitem{Choi:2017ylo}
	S.~Choi and R.~Akhoury, ``{BMS Supertranslation Symmetry Implies Faddeev-Kulish
		Amplitudes},'' \href{http://dx.doi.org/10.1007/JHEP02(2018)171}{{\em JHEP}
		{\bf 02} (2018)  171}, \href{http://arxiv.org/abs/1712.04551}{{\tt
			arXiv:1712.04551 [hep-th]}}.
	
	\bibitem{Hannesdottir:2019opa}
	H.~Hannesdottir and M.~D. Schwartz, ``{$S$ -Matrix for massless particles},''
	\href{http://dx.doi.org/10.1103/PhysRevD.101.105001}{{\em Phys. Rev. D} {\bf
			101} (2020) no.~10, 105001}, \href{http://arxiv.org/abs/1911.06821}{{\tt
			arXiv:1911.06821 [hep-th]}}.
	
	\bibitem{Damour:2020tta}
	T.~Damour, ``{Radiative contribution to classical gravitational scattering at
		the third order in $G$},''
	\href{http://dx.doi.org/10.1103/PhysRevD.102.124008}{{\em Phys. Rev. D} {\bf
			102} (2020) no.~12, 124008}, \href{http://arxiv.org/abs/2010.01641}{{\tt
			arXiv:2010.01641 [gr-qc]}}.
	
	\bibitem{Veneziano:2022zwh}
	G.~Veneziano and G.~A. Vilkovisky, ``{Angular momentum loss in gravitational
		scattering, radiation reaction, and the Bondi gauge ambiguity},''
	\href{http://dx.doi.org/10.1016/j.physletb.2022.137419}{{\em Phys. Lett. B}
		{\bf 834} (2022)  137419}, \href{http://arxiv.org/abs/2201.11607}{{\tt
			arXiv:2201.11607 [gr-qc]}}.
	
	\bibitem{Javadinezhad:2022ldc}
	R.~Javadinezhad and M.~Porrati, ``{Supertranslation-Invariant Formula for the
		Angular Momentum Flux in Gravitational Scattering},''
	\href{http://dx.doi.org/10.1103/PhysRevLett.130.011401}{{\em Phys. Rev.
			Lett.} {\bf 130} (2023) no.~1, 011401},
	\href{http://arxiv.org/abs/2211.06538}{{\tt arXiv:2211.06538 [gr-qc]}}.
	
	\bibitem{Riva:2023xxm}
	M.~M. Riva, F.~Vernizzi, and L.~K. Wong, ``{Angular momentum balance in
		gravitational two-body scattering: Flux, memory, and supertranslation
		invariance},'' \href{http://arxiv.org/abs/2302.09065}{{\tt arXiv:2302.09065
			[gr-qc]}}.
	
	\bibitem{Compere:2023qoa}
	G.~Comp\`ere, S.~E. Gralla, and H.~Wei, ``{An asymptotic framework for
		gravitational scattering},'' \href{http://arxiv.org/abs/2303.17124}{{\tt
			arXiv:2303.17124 [gr-qc]}}.
	
	\bibitem{Note3}
	To leading order, both the amplitude $\protect \mathcal A^{\mu \nu }$ \cite
	{Luna:2017dtq,DiVecchia:2021bdo,Cristofoli:2021vyo} and the stress-energy
	tensor obtained from worldline methods $t^{\mu \nu }$ \cite
	{Mougiakakos:2021ckm,Riva:2021vnj,Mougiakakos:2022sic} are directly linked to
	the classical waveform and, from this common quantity, one can then easily
	fix the relative factor. This simple connection is only valid to leading
	order (tree level), since appropriate subtractions are needed to obtain the
	subleading waveform from the one-loop amplitude \cite
	{Brandhuber:2023hhy,Herderschee:2023fxh,Georgoudis:2023lgf}. See \protect
	\eqref {Aq1q2k} for our conventions on $q_1$, $q_2$ and $k$. As anticipated,
	we drop static $\delta (\omega )$ contributions in $\protect \mathcal
	{A}^{\mu \nu }$.
	
	\bibitem{Note4}
	See \cite {Bern:2020uwk,Cheung:2020sdj} for the mapping between field-basis and
	worldline-basis tidal operators.
	
	\bibitem{Blanchet:2013haa}
	L.~Blanchet, ``{Gravitational Radiation from Post-Newtonian Sources and
		Inspiralling Compact Binaries},''
	\href{http://dx.doi.org/10.12942/lrr-2014-2}{{\em Living Rev. Rel.} {\bf 17}
		(2014)  2}, \href{http://arxiv.org/abs/1310.1528}{{\tt arXiv:1310.1528
			[gr-qc]}}.
	
	\bibitem{LIGOScientific:2017vwq}
	{\bf LIGO Scientific, Virgo} Collaboration, B.~P. Abbott {\em et al.},
	``{GW170817: Observation of Gravitational Waves from a Binary Neutron Star
		Inspiral},'' \href{http://dx.doi.org/10.1103/PhysRevLett.119.161101}{{\em
			Phys. Rev. Lett.} {\bf 119} (2017) no.~16, 161101},
	\href{http://arxiv.org/abs/1710.05832}{{\tt arXiv:1710.05832 [gr-qc]}}.
	
	\bibitem{Note5}
	With reference to Eq.~\protect \eqref {Jcomplete}, the fraction of the initial
	angular momentum that is lost due to tidal effects scales like $J_\protect
	\text {tid}/J\sim (Gm/b)^7 k_1^{(2)}/K_1^5$ while the fraction lost due to
	point-particle effects scales like $J_\protect \text {pp}/J\sim (Gm/b)^3$
	\cite {Manohar:2022dea}.
	
	\bibitem{Luna:2017dtq}
	A.~Luna, I.~Nicholson, D.~O'Connell, and C.~D. White, ``{Inelastic Black Hole
		Scattering from Charged Scalar Amplitudes},''
	\href{http://dx.doi.org/10.1007/JHEP03(2018)044}{{\em JHEP} {\bf 03} (2018)
		044}, \href{http://arxiv.org/abs/1711.03901}{{\tt arXiv:1711.03901
			[hep-th]}}.
	
	\bibitem{Lee:2012cn}
	R.~N. Lee, ``{Presenting LiteRed: a tool for the Loop InTEgrals REDuction},''
	\href{http://arxiv.org/abs/1212.2685}{{\tt arXiv:1212.2685 [hep-ph]}}.
	
	\bibitem{Lee:2013mka}
	R.~N. Lee, ``{LiteRed 1.4: a powerful tool for reduction of multiloop
		integrals},'' \href{http://dx.doi.org/10.1088/1742-6596/523/1/012059}{{\em J.
			Phys. Conf. Ser.} {\bf 523} (2014)  012059},
	\href{http://arxiv.org/abs/1310.1145}{{\tt arXiv:1310.1145 [hep-ph]}}.
	
	\bibitem{DEath:1976bbo}
	P.~D. D'Eath, ``{High Speed Black Hole Encounters and Gravitational
		Radiation},''
	\href{http://dx.doi.org/10.1103/PhysRevD.18.990}{{\em Phys. Rev.} {\bf D18}
		(1978)  990}.
	%%CITATION = PHRVA,D18,990;%%.
	
	\bibitem{Kovacs:1977uw}
	S.~J. Kovacs and K.~S. Thorne, ``{The Generation of Gravitational Waves. 3.
		Derivation of Bremsstrahlung Formulas},''
	\href{http://dx.doi.org/10.1086/155576}{{\em Astrophys. J.} {\bf 217} (1977)
		252--280}.
	%%CITATION = ASJOA,217,252;%%.
	
	\bibitem{Kovacs:1978eu}
	S.~J. Kovacs and K.~S. Thorne, ``{The Generation of Gravitational Waves. 4.
		Bremsstrahlung},''
	\href{http://dx.doi.org/10.1086/156350}{{\em Astrophys. J.} {\bf 224} (1978)
		62--85}.
	%%CITATION = ASJOA,224,62;%%.
	
	\bibitem{Bini:2018ywr}
	D.~Bini and T.~Damour, ``{Gravitational spin-orbit coupling in binary systems
		at the second post-Minkowskian approximation},''
	\href{http://dx.doi.org/10.1103/PhysRevD.98.044036}{{\em Phys. Rev.} {\bf
			D98} (2018) no.~4, 044036},
	\href{http://arxiv.org/abs/1805.10809}{{\tt arXiv:1805.10809 [gr-qc]}}.
	%%CITATION = ARXIV:1805.10809;%%.
	
	\bibitem{Bini:2021gat}
	D.~Bini, T.~Damour, and A.~Geralico, ``{Radiative contributions to
		gravitational scattering},''
	\href{http://dx.doi.org/10.1103/PhysRevD.104.084031}{{\em Phys. Rev. D} {\bf
			104} (2021) no.~8, 084031}, \href{http://arxiv.org/abs/2107.08896}{{\tt
			arXiv:2107.08896 [gr-qc]}}.
	
	\bibitem{Bini:2012ji}
	D.~Bini and T.~Damour, ``{Gravitational radiation reaction along general orbits
		in the effective one-body formalism},''
	\href{http://dx.doi.org/10.1103/PhysRevD.86.124012}{{\em Phys. Rev. D} {\bf
			86} (2012)  124012}, \href{http://arxiv.org/abs/1210.2834}{{\tt
			arXiv:1210.2834 [gr-qc]}}.
	
	\bibitem{Kalin:2019rwq}
	G.~K{\"a}lin and R.~A. Porto, ``{From Boundary Data to Bound States},''
	\href{http://dx.doi.org/10.1007/JHEP01(2020)072}{{\em JHEP} {\bf 01} (2020)
		072}, \href{http://arxiv.org/abs/1910.03008}{{\tt arXiv:1910.03008
			[hep-th]}}.
	
	\bibitem{Kalin:2019inp}
	G.~K{\"a}lin and R.~A. Porto, ``{From boundary data to bound states. Part II.
		Scattering angle to dynamical invariants (with twist)},''
	\href{http://dx.doi.org/10.1007/JHEP02(2020)120}{{\em JHEP} {\bf 02} (2020)
		120}, \href{http://arxiv.org/abs/1911.09130}{{\tt arXiv:1911.09130
			[hep-th]}}.
	
	\bibitem{Saketh:2021sri}
	M.~V.~S. Saketh, J.~Vines, J.~Steinhoff, and A.~Buonanno, ``{Conservative and
		radiative dynamics in classical relativistic scattering and bound systems},''
	\href{http://dx.doi.org/10.1103/PhysRevResearch.4.013127}{{\em Phys. Rev.
			Res.} {\bf 4} (2022) no.~1, 013127},
	\href{http://arxiv.org/abs/2109.05994}{{\tt arXiv:2109.05994 [gr-qc]}}.
	
	\bibitem{Cho:2021arx}
	G.~Cho, G.~K\"alin, and R.~A. Porto, ``{From boundary data to bound states.
		Part III. Radiative effects},''
	\href{http://dx.doi.org/10.1007/JHEP04(2022)154}{{\em JHEP} {\bf 04} (2022)
		154}, \href{http://arxiv.org/abs/2112.03976}{{\tt arXiv:2112.03976
			[hep-th]}}.
	
	\bibitem{Peters:1964zz}
	P.~C. Peters, ``{Gravitational Radiation and the Motion of Two Point Masses},''
	\href{http://dx.doi.org/10.1103/PhysRev.136.B1224}{{\em Phys. Rev.} {\bf 136}
		(1964)  B1224--B1232}.
	
	\bibitem{Thorne:1980ru}
	K.~S. Thorne, ``{Multipole Expansions of Gravitational Radiation},''
	\href{http://dx.doi.org/10.1103/RevModPhys.52.299}{{\em Rev. Mod. Phys.} {\bf
			52} (1980)  299--339}.
	
	\bibitem{Damour:1981bh}
	T.~Damour and N.~Deruelle, ``{Radiation Reaction and Angular Momentum Loss in
		Small Angle Gravitational Scattering},''
	\href{http://dx.doi.org/10.1016/0375-9601(81)90567-3}{{\em Phys. Lett. A}
		{\bf 87} (1981)  81}.
	
	\bibitem{Bonga:2018gzr}
	B.~Bonga and E.~Poisson, ``{Coulombic contribution to angular momentum flux in
		general relativity},''
	\href{http://dx.doi.org/10.1103/PhysRevD.99.064024}{{\em Phys. Rev. D} {\bf
			99} (2019) no.~6, 064024}, \href{http://arxiv.org/abs/1808.01288}{{\tt
			arXiv:1808.01288 [gr-qc]}}.
	
	\bibitem{Blanchet:2018yqa}
	L.~Blanchet and G.~Faye, ``{Flux-balance equations for linear momentum and
		center-of-mass position of self-gravitating post-Newtonian systems},''
	\href{http://dx.doi.org/10.1088/1361-6382/ab0d4f}{{\em Class. Quant. Grav.}
		{\bf 36} (2019) no.~8, 085003}, \href{http://arxiv.org/abs/1811.08966}{{\tt
			arXiv:1811.08966 [gr-qc]}}.
	
	\bibitem{AbhishekChowdhuri:2022ora}
	A.~Chowdhuri and A.~Bhattacharyya, ``{Study of eccentric binaries in Horndeski
		gravity},'' \href{http://dx.doi.org/10.1103/PhysRevD.106.064046}{{\em Phys.
			Rev. D} {\bf 106} (2022) no.~6, 064046},
	\href{http://arxiv.org/abs/2203.09917}{{\tt arXiv:2203.09917 [gr-qc]}}.
	
	\bibitem{Note6}
	This can be obtained by first noting that $\protect \dot r = v^2_\protect \text
	{rel} t/r= v_\protect \text {rel}\protect \sqrt {1-(b/r)^2}$, and then
	performing the change of variable $r = b/\sin \theta $.
	
	\bibitem{Riva:2022fru}
	M.~M. Riva, F.~Vernizzi, and L.~K. Wong, ``{Gravitational bremsstrahlung from
		spinning binaries in the post-Minkowskian expansion},''
	\href{http://dx.doi.org/10.1103/PhysRevD.106.044013}{{\em Phys. Rev. D} {\bf
			106} (2022) no.~4, 044013}, \href{http://arxiv.org/abs/2205.15295}{{\tt
			arXiv:2205.15295 [hep-th]}}.
	
	\bibitem{Henry:2019xhg}
	Q.~Henry, G.~Faye, and L.~Blanchet, ``{Tidal effects in the equations of motion
		of compact binary systems to next-to-next-to-leading post-Newtonian order},''
	\href{http://dx.doi.org/10.1103/PhysRevD.101.064047}{{\em Phys. Rev. D} {\bf
			101} (2020) no.~6, 064047}, \href{http://arxiv.org/abs/1912.01920}{{\tt
			arXiv:1912.01920 [gr-qc]}}.
	
	\bibitem{Henry:2020ski}
	Q.~Henry, G.~Faye, and L.~Blanchet, ``{Tidal effects in the gravitational-wave
		phase evolution of compact binary systems to next-to-next-to-leading
		post-Newtonian order},''
	\href{http://dx.doi.org/10.1103/PhysRevD.102.044033}{{\em Phys. Rev. D} {\bf
			102} (2020) no.~4, 044033}, \href{http://arxiv.org/abs/2005.13367}{{\tt
			arXiv:2005.13367 [gr-qc]}}.
	
	\bibitem{Henry:2020pzq}
	Q.~Henry, G.~Faye, and L.~Blanchet, ``{Hamiltonian for tidal interactions in
		compact binary systems to next-to-next-to-leading post-Newtonian order},''
	\href{http://dx.doi.org/10.1103/PhysRevD.102.124074}{{\em Phys. Rev. D} {\bf
			102} (2020) no.~12, 124074}, \href{http://arxiv.org/abs/2009.12332}{{\tt
			arXiv:2009.12332 [gr-qc]}}.
	
	\bibitem{Brandhuber:2023hhy}
	A.~Brandhuber, G.~R. Brown, G.~Chen, S.~De~Angelis, J.~Gowdy, and
	G.~Travaglini, ``{One-loop gravitational bremsstrahlung and waveforms from a
		heavy-mass effective field theory},''
	\href{http://dx.doi.org/10.1007/JHEP06(2023)048}{{\em JHEP} {\bf 06} (2023)
		048}, \href{http://arxiv.org/abs/2303.06111}{{\tt arXiv:2303.06111
			[hep-th]}}.
	
	\bibitem{Herderschee:2023fxh}
	A.~Herderschee, R.~Roiban, and F.~Teng, ``{The sub-leading scattering waveform
		from amplitudes},'' \href{http://dx.doi.org/10.1007/JHEP06(2023)004}{{\em
			JHEP} {\bf 06} (2023)  004}, \href{http://arxiv.org/abs/2303.06112}{{\tt
			arXiv:2303.06112 [hep-th]}}.
	
	\bibitem{Georgoudis:2023lgf}
	A.~Georgoudis, C.~Heissenberg, and I.~Vazquez-Holm, ``{Inelastic exponentiation
		and classical gravitational scattering at one loop},''
	\href{http://dx.doi.org/10.1007/JHEP06(2023)126}{{\em JHEP} {\bf 06} (2023)
		126}, \href{http://arxiv.org/abs/2303.07006}{{\tt arXiv:2303.07006
			[hep-th]}}.
	
\end{thebibliography}
\end{document}